\documentclass[11pt,prd,onecolumn,amsmath,amssymb,aps,floats,floatfix,nofootinbib]{revtex4-2}
\usepackage[colorlinks=true,urlcolor=blue,anchorcolor=blue,citecolor=blue,filecolor=blue,linkcolor=blue,menucolor=blue,linktocpage=true]{hyperref} 

\usepackage[inline]{enumitem}
\usepackage[multidot]{grffile}  
\usepackage{dcolumn}
\usepackage{bm}
\usepackage{amsmath}
\usepackage{amsfonts}
\usepackage{amssymb}
\usepackage{tikz}
\usetikzlibrary{positioning}
\usetikzlibrary{decorations.pathreplacing}
\usetikzlibrary{decorations.markings}
\usepackage{color}
\usepackage{float}
\usepackage{latexsym}
\usepackage{slashed} 
\usepackage{pstricks}
\usepackage{indentfirst}

\usepackage{mathrsfs}
\usepackage{multirow}
\usepackage{epsfig,psfrag}
\usepackage{graphicx}
\usepackage{enumitem}
\usepackage{geometry,amssymb,yfonts}
\usepackage{yhmath}
\usepackage{anysize}
\usepackage{subfigure}
\usepackage{mathtools}
\usepackage{setspace} 
\usepackage[utf8]{inputenc} 
\usepackage[scientific-notation=true]{siunitx} 
\usepackage{siunitx}
\usepackage{url}
\usepackage{makecell}

\graphicspath{{fig/}}

\allowdisplaybreaks 

\newcommand{\lagrange}{\ensuremath{\mathcal{L}}}

\newcommand{\UoneY}{\mathrm{U}(1)_\mathrm{Y}}
\newcommand{\SUtwoL}{\mathrm{SU}(2)_\mathrm{L}}

\begin{document}
	
	\title{Leptogenesis assisted by scalar decays}
	
	\author{Jun-Yu Tong}
	\author{Zhao-Huan Yu}\email[Corresponding author. ]{yuzhaoh5@mail.sysu.edu.cn}
	\author{Hong-Hao Zhang}\email[Corresponding author. ]{zhh98@mail.sysu.edu.cn}
	\affiliation{School of Physics, Sun Yat-Sen University, Guangzhou 510275, China}
	
	\begin{abstract}
    We present a pragmatic approach to lower down the mass scale of right-handed neutrinos in leptogenesis by introducing a scalar decaying to right-handed neutrinos. The key point of our proposal is that the out-of-equilibrium decays of the scalar provide an additional source for right-handed neutrinos and hence the lepton asymmetry. This mechanism works well at low temperatures when the washout of the generated lepton asymmetry is suppressed. Thus, the lepton asymmetry can be effectively produced despite the washout effect is strong or not. Through a comprehensive analysis, we demonstrate that such a scalar-assisted leptogenesis can typically decrease the viable right-handed neutrino mass scale by one to three orders of magnitude.
	\end{abstract}
	
	\maketitle
	\tableofcontents
	
	\clearpage
	
	\section{Introduction}
	
	Although matter and antimatter are treated symmetrically in particle physics and quantum field theory, stars and galaxies in the celestial neighborhood are constituted exclusively by baryonic matter.
	Actually, a matter-antimatter symmetric observable universe is excluded by the observation of the cosmic diffuse $\gamma$-ray background~\cite{Cohen:1997ac}.
	The baryon asymmetry of the universe is conventionally quantified by the net baryon-to-photon number density $\eta_B = (n_B - n_{\bar{B}})/n_\gamma$.
	Combining the observations of the cosmic microwave background by the Planck collaboration~\cite{Planck:2018nkj} and the primordial abundances of light elements related to big bang nucleosynthesis, $\eta_B$ is determined to be~\cite{Fields:2019pfx}
	\begin{equation}\label{eq:eta_B_obs}
		\eta_B = (6.13 \pm 0.04) \times 10^{-10}.
	\end{equation}
	
	In order to generate such a baryon asymmetry, the related physical processes must satisfy three Sakharov conditions~\cite{Sakharov:1967dj}, i.e., the  violation of the baryon number $B$, the $C$ and $CP$ violations, and the departure from thermal equilibrium.
	The standard model (SM) by itself could fulfill all the three conditions.
	The $B$ violation occurs through a nonperturbative effect in electroweak interactions known as the sphaleron processes~\cite{tHooft:1976rip, Rubakov:1996vz}, and the interactions out of thermal equilibrium may happen in a strong first-order electroweak phase transition.
	This makes electroweak baryogenesis~\cite{Kuzmin:1985mm} a promising mechanism to generate the baryon asymmetry.
	However, subsequent researches reveal that the electroweak phase transition in the SM is just a smooth crossover because the Higgs boson mass is too large~\cite{Buchmuller:1994qy, Kajantie:1996mn, Csikor:1998eu}.
	Thus, a successful baryogenesis can only take place in new physics beyond the SM.
	Some comprehensive reviews can be found in Refs.~\cite{Dine:2003ax,Bodeker:2020ghk,Pereira:2023xiw}.

	Among new physics theories trying to explain the origin of the baryon asymmetry,
	leptogenesis~\cite{Fukugita:1986hr, Luty:1992un} is a well-motivated mechanism  related to the tiny masses of active neutrinos.
	It typically introduces three generations of heavy right-handed neutrinos (RHNs), allowing active neutrinos to acquire Majorana masses through the type-I seesaw mechanism~\cite{Minkowski:1977sc, Gell-Mann:1979vob, Yanagida:1979as}.
	The decays of RHNs give rise to the lepton asymmetry, which is subsequently converted into the baryon asymmetry via the sphaleron processes.
	Such a leptogenesis mechanism can be realized in various models (see Refs.~\cite{Kusenko:2014uta, Pearce:2015nga, Pascoli:2016gkf, Agashe:2018oyk, Barrie:2021mwi, Berman:2022oht, Bhattacharya:2021jli, Fernandez-Martinez:2022gsu, Carrasco-Martinez:2023nit, Datta:2021gyi, Co:2019wyp, Zhao:2020bzx, Enomoto:2020lpf, Jiang:2020kbt, DuttaBanik:2020vfr, Granelli:2023egb,Chen:2024arl} for an incomplete list of recent works).
	
Nevertheless, the standard thermal leptogenesis typically requires a high mass scale of the RHNs.
For achieving an adequate $CP$ violation in RHN decays, the lightest RHN mass should be larger than $10^9~\si{GeV}$~\cite{Davidson:2002qv}.
Moreover, the generated lepton asymmetry would be strongly washed out if the RHN Yukawa couplings are too large~\cite{Buchmuller:2002rq, Buchmuller:2002jk, Buchmuller:2003gz}.
Taking into account the washout effect and the neutrino oscillation data, it demands the lightest RHN mass to be at least $\sim 10^{10}~\si{GeV}$ for obtaining the correct baryon-to-photon ratio~\cite{Buchmuller:2002rq}.
Such a high mass scale is challenging to be further tested by experiments.

	In order to circumvent a strong washout effect and/or lower down the RHN mass scale, several solutions were proposed.
	One solution is to assume leptogenesis originating from the reheating process after inflation~\cite{Asaka:1999yd, Hahn-Woernle:2008tsk, Barman:2021tgt}.
	Another approach is to reprocess $CP$-violating interactions at high scales into the $B-L$ asymmetry by RHN interactions at lower scales, known as wash-in leptogenesis~\cite{Domcke:2020quw, Domcke:2022kfs}.
	Additional avenues include considering nonthermal particle production from the evaporation of primordial black holes~\cite{Datta:2020bht, Barman:2021ost, Calabrese:2023key, Schmitz:2023pfy, Ghoshal:2023fno}, imposing restrictions on the related couplings~\cite{Suematsu:2019kst, Huang:2023gse}, embedding leptogenesis within a first-order phase transition~\cite{Chun:2023ezg}, and assuming nonstandard cosmology~\cite{Dehpour:2023dfo, Dehpour:2023wyy}.
	Besides, a lower RHN mass scale can also be achieved by resonant leptogenesis~\cite{Pilaftsis:2003gt, Pilaftsis:2004xx, Dev:2017wwc, daSilva:2022mrx, Karamitros:2023tqr, Zhao:2024uid}, flavored leptogenesis~\cite{Barbieri:1999ma, Abada:2006ea, Moffat:2018wke}, and extra $CP$-violation by an additional scalar~\cite{Alanne:2018brf}.
 
	Motivated by these efforts, we propose another approach to decrease the RHN mass scale by introducing a scalar boson $\phi$ that decays into RHNs when the $\phi$ particles are out of thermal equilibrium\footnote{The impact of a scalar coupling to RHNs has also been studied in Refs.~\cite{Dev:2017xry, Barreiros:2022fpi} for other motivations.}.
	We assume a slow decay rate of $\phi$ to ensure the continuous production of RHNs, even when the washout processes are suppressed at low temperatures.
	Thus, it provides an additional nonthermal source for the lepton asymmetry.
	By tracking the particle number density evolution in this scenario, we will explore the suitable parameter regions that can effectively lower down the lightest RHN mass.
	
	The paper is structured as follows.
	In Section~\ref{sec:Model}, we introduce our model, which extends the SM by incorporating a scalar field and three RHNs, and outline the calculations of the $CP$ asymmetry and the requisite decay widths and scattering cross sections.
	Section~\ref{sec:Boltzmann} is dedicated to analyzing the Boltzmann equations and number densities to assess the efficacy of the model.
	In Section~\ref{sec:scan}, we explore the viable parameter regions by performing parameter scans.
	Finally, we conclude this work in Section~\ref{sec:sum}.
	
	\section{Model}
	\label{sec:Model}
	
	We extend the SM with three generations of RHNs $N_{i\mathrm{R}}$ ($i=1,2,3$) and a real scalar $\phi$, which is an SM gauge singlet. The $CP$-violating decays of the RHNs are expected to give rise to an asymmetry in the lepton number $L$.
	The related Lagrangian is
	\begin{eqnarray}
		-\lagrange &\supset& \left( h_{\nu,{ij}}\overline{L_{i\mathrm{L}}}\tilde{H}N_{j\mathrm{R}}+\frac{1}{2}M_{N_i}\overline{N^\mathrm{c}_{i\mathrm{R}}}N_{i\mathrm{R}}+\frac{1}{2}y_i\overline{N^\mathrm{c}_{i\mathrm{R}}}N_{i\mathrm{R}}\phi + \text{H.c.}\right)
		\nonumber\\
		&&	+\, m_\phi^2\phi^2 + \kappa H^\dagger H\phi+ \lambda_{\phi H } H^\dagger H\phi^2.
	\end{eqnarray}
	Here $L_{i\mathrm{L}}$ represent the $\SUtwoL$ doublets of the left-handed leptons, and $\tilde{H}=i\sigma_2 H^*$ is the iso-doublet of the SM Higgs field
	\begin{equation}
		H = \begin{pmatrix}
			H^+\\
			H^0
		\end{pmatrix},
	\end{equation}
	which develops a vacuum expectation value $v = 174~\si{GeV}$, leading to the spontaneous breaking of the electroweak gauge symmetry.
	$N_{i\mathrm{R}}$ carry no $\UoneY$ charges and weakly couple to the real scalar $\phi$ via the Yukawa couplings $y_i$.
        $N_{i\mathrm{R}}^\mathrm{c}$ are the charge conjugates of $N_{i\mathrm{R}}$.
	The Yukawa couplings $h_{\nu,ij}$ give Dirac mass terms for the neutrinos after electroweak symmetry breaking, while $M_{N_i}$ provide Majorana mass terms.
	Among the couplings between $\phi$ and the SM Higgs doublet $H$, $\kappa$ is a dimension-1 coupling constant, while $\lambda_{\phi H}$ is dimensionless.
	We assume that $\phi$ does not develop a nonzero vacuum expectation value and its mass is just given by $m_\phi$.
	
	After the electroweak symmetry breaking, the mass terms for the neutrinos become
	\begin{equation}
		-\lagrange_{\mathrm{mass}} = \frac{1}{2} 
		\begin{pmatrix}
			\overline{\nu_\mathrm{L}} &
			\overline{N^\mathrm{c}_{\mathrm{R}}}
		\end{pmatrix}
		\begin{pmatrix}
			0 & vh_\nu\\
			vh_\nu^\mathrm{T} & M
		\end{pmatrix}
		\begin{pmatrix}
			\nu_\mathrm{L}^\mathrm{c} \\
			N_{\mathrm{R}}
		\end{pmatrix},
	\end{equation}
	where $M=\operatorname{diag}(M_{N_1},M_{N_2},M_{N_3})$ is the diagonal mass matrix for RHNs.
	Through a block-diagonalization of the $6\times 6$ mass matrix in $\lagrange_{\mathrm{mass}}$, the $3\times 3$ mass matrix for the active neutrinos is given by
	\begin{equation}\label{3}
		M_\nu \simeq -v^2 h_\nu {M}^{-1} h_\nu^\mathrm{T}
	\end{equation}
	for $M \gg v h_\nu$.
	Thus, the large RHN masses give an origin to the tiny masses of the active neutrinos via the type-I seesaw mechanism.
	On the other hand, the flavor eigenstates of the RHNs basically coincide with their mass eigenstates, and three Majorana spinor fields can be constructed by $N_i = N_{i\mathrm{R}}^\mathrm{c} + N_{i\mathrm{R}}$.
	The masses of the corresponding heavy Majorana neutrinos $N_i$ are approximately given by $M_{N_i}$.
	
	A specific form of the Yukawa coupling matrix $h_\nu$ can be derived by the Casas-Ibarra parametrization~\cite{Casas:2001sr}.
	The active neutrino mass matrix $M_\nu$, which is complex and symmetric, can be diagonalized by the unitary Pontecorvo–Maki–Nakagawa–Sakata (PMNS) matrix $U$~\cite{Pontecorvo:1957qd, Maki:1962mu}:
	\begin{equation}\label{4}
		U^\mathrm{T} M_\nu U = M_\nu^\mathrm{d},
	\end{equation}
	where $M_\nu^\mathrm{d} = \operatorname{diag}(m_1, m_2, m_3)$.
	Blending Eqs.~\eqref{3} and \eqref{4}, we have
	\begin{equation}
		-v^2 U^\mathrm{T} h_\nu M^{-1} h_\nu^\mathrm{T} U=M_\nu^\mathrm{d},
	\end{equation}
	and it is not difficult to obtain
	\begin{equation}
		-v^2\left[\left(\sqrt{M_\nu^\mathrm{d}}\right)^{-1} U^\mathrm{T} h_\nu \left(\sqrt{M}\right)^{-1} \right]\left[\left(\sqrt{M}\right)^{-1} h_\nu^\mathrm{T} U \left(\sqrt{M_\nu^\mathrm{d}}\right)^{-1} \right]=I_{3\times 3},
	\end{equation}
	where the square root of a diagonal matrix means the square roots of its diagonal elements.
	Thus,
	\begin{equation}
		R^\mathrm{T} \equiv -i v\left(\sqrt{M_\nu^\mathrm{d}}\right)^{-1} U^\mathrm{T} h_\nu \left(\sqrt{M}\right)^{-1}
	\end{equation}
	must be a $3 \times 3$ complex orthogonal matrix satisfying $R^\mathrm{T} R = I_{3\times 3}$.
	Therefore, the Yukawa coupling matrix can be expressed as
	\begin{equation}\label{eq:h_nu}
		h_\nu=\frac{i}{v}\, U^*\sqrt{M^\mathrm{d}_\nu}R^\mathrm{T}\sqrt{M},
	\end{equation}
    which implies that larger values in the RHN mass matrix $M$ leads to larger Yukawa couplings $h_\nu$ if the other parameters are fixed.
	
	Following Particle Data Group's convention~\cite{Workman:2022ynf} for the parametrization of the PMNS matrix, we have
	\begin{equation}
		U=\left(\begin{matrix}
			c_{12}c_{13}&s_{12}c_{13}&s_{13}e^{-i\delta_{CP}}\\
			-s_{12}c_{23}-c_{12}s_{13}s_{23}e^{i\delta_{CP}}&c_{12}c_{23}-s_{12}s_{13}s_{23}e^{i\delta_{CP}}&c_{13}s_{23}\\
			s_{12}s_{23}-c_{12}s_{13}c_{23}e^{i\delta_{CP}}&-c_{12}s_{23}-s_{12}s_{13}c_{23}e^{i\delta_{CP}}&c_{13}c_{23}
		\end{matrix}\right) \left(\begin{matrix}
			e^{i\eta_1}&  0&0\\
			0&e^{i\eta_2}&0\\
			0&0&1
		\end{matrix}\right),
	\end{equation}
	where $c_{ij}\equiv \cos{\theta_{ij}}$, $ s_{ij}\equiv \sin{\theta_{ij}}$.
	$\delta_{CP}$ is the Dirac $CP$ phase and $\eta_{1,2}$ are the Majorana $CP$ phases.
	In the following calculation, we assume a normal hierarchy for the active neutrino masses, and  set the Majorana phases to be zero and the neutrino oscillation parameters to be the best fit values from the global analysis given by the NuFIT collaboration~\cite{Gonzalez-Garcia:2021dve}, which are listed in Table~\ref{tab:bestfit}.

	\begin{table}[!t]
		\centering
		\setlength\tabcolsep{.7em}
		\renewcommand{\arraystretch}{1.5}
		\caption{Three-flavor global fit results of  neutrino oscillation parameters in $1\sigma$ range.}\label{tab:bestfit}
		\begin{tabular}{ccccccc}
			\hline\hline
			$\theta_{12}~(\si{\degree})$ &  $\theta_{23}~(\si{\degree})$    &  $\theta_{13}~(\si{\degree})$   &  $\delta_{CP}~(\si{\degree})$ &  $\Delta m_{21}^2~(10^{-5}~\si{eV}^2)$ & $\Delta m_{31}^2~(10^{-3}~\si{eV}^2)$ \\
			\hline
			$33.44^{+0.77}_{-0.74}$&$49.2^{+1.0}_{-1.3}$&$8.57^{+0.13}_{-0.12}$&$194^{+52}_{-25}$&$7.42^{+0.21}_{-0.20}$&$2.515^{+0.028}_{-0.028}$\\
			\hline\hline
		\end{tabular}
	\end{table}
	
	Moreover, we assume the lightest neutrino mass $m_1$ to be zero, leading to $m_2 = \sqrt{\Delta m_{21}^2}$ and $m_3 = \sqrt{\Delta m_{31}^2}$.
	In this case, the complex orthogonal matrix $R$ can be described by a single complex parameter $z = z_\mathrm{r} + iz_\mathrm{i}$~\cite{Antusch:2011nz} with a form of
	\begin{equation}
		R=\begin{pmatrix}
			0&\cos z&\sin z\\
			0&-\sin z&\cos z\\
			1&0&0
		\end{pmatrix}.
	\end{equation}
	for the normal hierarchy.
	Thus, given the values of $m_{N_1}$, $m_{N_2}$, $m_{N_3}$, $z_\mathrm{r}$, and $z_\mathrm{i}$, the Yukawa coupling matrix $h_\nu$ can be determined through Eq.~\eqref{eq:h_nu}.
	
	The $CP$ asymmetry in the decays of the heavy Majorana neutrino $N_i$ to light leptons and Higgs bosons can be measured by
	\begin{equation}
		\varepsilon_i=\frac{\Gamma(N_i\to \ell H)-\Gamma(N_i\to \bar{\ell} \bar{H})}{\Gamma(N_i\to \ell H)+\Gamma(N_i\to \bar{\ell} \bar{H})},
	\end{equation}
	where $\ell = \ell_j^-, \nu_j~(j=1,2,3)$ and $H = H^+, H^0$, while $\bar\ell$ and $\bar{H}$ represent their antiparticles.
	Considering the interference of one-loop diagrams with tree-level diagrams, it is given by~\cite{Covi:1996wh}
	\begin{equation}
		\varepsilon_i=\frac{1}{8\pi}\sum_{j\neq i}\left[f\left(\frac{M_{N_j}^2}{M_{N_i}^2}\right)-\frac{M_{N_i} M_{N_j}}{M_{N_j}^2-M_{N_i}^2}\right]\frac{\operatorname{Im}\{[(h_\nu^\dagger h_\nu)_{ij}]^2\}}{(h_\nu^\dagger h_\nu)_{ii}},
	\end{equation}
	where $f(x)=\sqrt{x}\{1-(1+x)\ln[(1+x)/x]\}$.
    Basically, $\varepsilon_i$ are positively correlated to the Yukawa couplings $h_{\nu,ij}$, which are positively correlated to $M_{N_i}$ as mentioned above.
	When the $CP$-violating $N_i$ decays are out of thermal equilibrium, a lepton number asymmetry would be generated.
	
\begin{figure}[!t]
		\centering
		\includegraphics[width=0.5\linewidth]{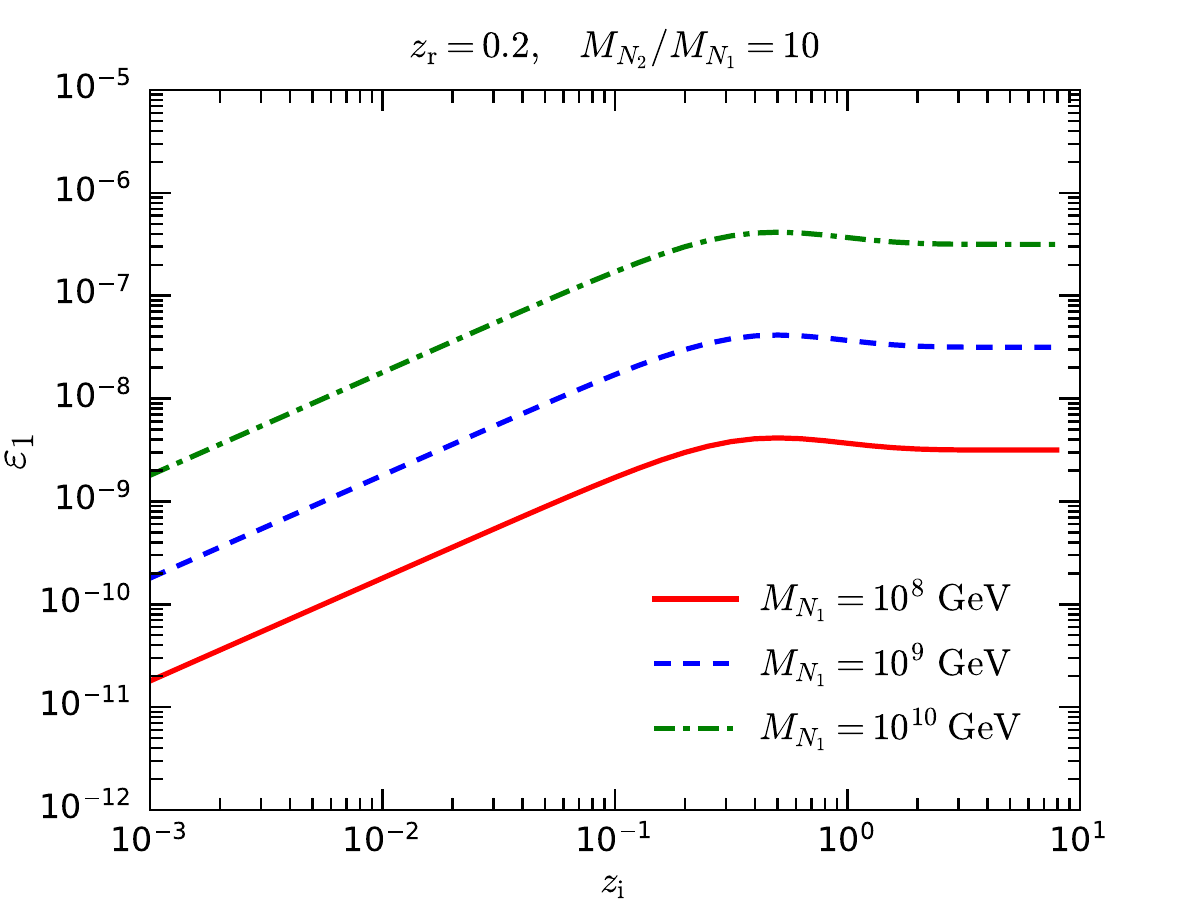}
		\caption{The $CP$ asymmetry  $\varepsilon_1$ as functions of $z_\mathrm{i}$ for $M_{N_1} = 10^8, 10^9, $ and $10^{10}~\si{GeV}$ with $z_\mathrm{r} = 0.2$, $M_{N_2} = 10 M_{N_1}$, and $M_{N_3} \to \infty$.}
		\label{eps}
	\end{figure}

	In Fig.~\ref{eps}, we demonstrate the $CP$ asymmetry in the $N_1$ decays as functions of $z_\mathrm{i}$ for $M_{N_1} = 10^8, 10^9, $ and $10^{10}~\si{GeV}$.
	The other parameters are fixed as $z_\mathrm{r} = 0.2$, $M_{N_2} = 10 M_{N_1}$, and $M_{N_3} \to \infty$.
	It is evident that the $CP$ asymmetry $\varepsilon_1$ is basically proportional to $M_{N_1}$.
	Moreover, $\varepsilon_1$ increases as $z_\mathrm{i}$ increases for $z_\mathrm{i} \lesssim 0.4$, but it tends to be saturated for $z_\mathrm{i} \gtrsim 0.4$.
	
	For simplicity, we further assume a mass hierarchy of the heavy Majorana neutrinos as $M_{N_1} \ll M_{N_2}, M_{N_3}$.
	As a result, the lepton asymmetry generated by $N_2$ and $N_3$ will be washed out by the inverse decay processes $\ell H/\bar{\ell}\bar{H} \to N_1$.
	Therefore, only the $B-L$ numbers yielded from $N_1$ remain and are subsequently converted into the baryon asymmetry through the electroweak sphaleron processes~\cite{Rubakov:1996vz}.
	This is the one-flavor approximation.
	After the electroweak crossover, the sphaleron processes freeze out, and the resulting baryon-to-photon ratio at recombination can be estimated as~\cite{Bodeker:2020ghk}
	\begin{equation}
		\eta_B=\frac{c_s}{f}\frac{n_{B-L}}{n_\gamma}\simeq 0.013\, \frac{n_{B-L}}{n_\gamma}.
	\end{equation}
	Here, $n_{B-L}$ and $n_\gamma$ represent the number densities of the $B-L$ charges and the photons, respectively.
	$c_s = 28/79$ is the conversion ratio of the $B-L$ asymmetry to the baryon asymmetry via the sphaleron processes for the inclusion of three RHNs~\cite{Khlebnikov:1988sr}.
	$f = 2387/86$ is a dilution factor accounting for the increase in $n_\gamma$ in a comoving volume from the epoch with $T \sim m_{N_1}$ to recombination.
	
	
	The washout effect plays a crucial role in diminishing the final yield of $n_{B-L}$.
	The primary washout processes are the inverse decays $\ell H / \bar{\ell}\bar{H} \to N_1$.
	They are related to the decay width of $N_1$, which is given by
	\begin{equation}
		\Gamma_{N_1} = \Gamma(N_1\to \ell H)+\Gamma(N_1\to \bar{\ell}  \bar{H}) \simeq \frac{1}{8\pi}(h_\nu^\dagger h_\nu)_{11} M_{N_1},
	\end{equation}
	where the masses of light leptons and Higgs bosons have been neglected.
	
\begin{figure}[!t]
		\centering
		\begin{subfigure}
			\centering
			\begin{tikzpicture}
				\draw[dashed] (-1,0) node[left] {\(\phi\)} -- (0,0);
				\draw[dashed, postaction={decorate, decoration={
						markings,
						mark=at position 0.5 with {\arrow{>}}}
				}] (0,0) -- (1,1) node[right] {\(H^0~(H^+)\)};
				\draw[dashed, postaction={decorate, decoration={
						markings,
						mark=at position 0.5 with {\arrow{<}}}
				}] (0,0) -- (1,-1) node[right] {\(\bar{H}^0~(H^-)\)};
			\end{tikzpicture}
		\end{subfigure}
		\hspace{1.2em}
		\begin{subfigure}
			\centering
			\begin{tikzpicture}
				\draw[dashed] (-1,0) node[left] {\(\phi\)} -- (0,0);
				\draw (0,0) -- (1,1) node[right] {\(N_1\)};
				\draw (0,0) -- (1,-1) node[right] {\(N_1\)};
			\end{tikzpicture}
		\end{subfigure}
		\hspace{1.2em}
		\begin{subfigure}
			\centering
			\begin{tikzpicture}
				\draw[dashed] (-1,1) node[left] {\(\phi\)} -- (0,0);
				\draw[dashed] (-1,-1) node[left] {\(\phi\)} -- (0,0);
				\draw[dashed,postaction={decorate, decoration={
						markings,
						mark=at position 0.5 with {\arrow{>}}}
				}] (0,0) -- (1,1) node[right] {\(H^0~(H^+)\)};
				\draw[dashed,postaction={decorate, decoration={
						markings,
						mark=at position 0.5 with {\arrow{<}}}
				}] (0,0) -- (1,-1) node[right] {\(\bar{H}^0~(H^-)\)};
			\end{tikzpicture}
		\end{subfigure}
		\caption{Decay and scattering diagrams for the scalar $\phi$.}
		\label{feynman}
	\end{figure}
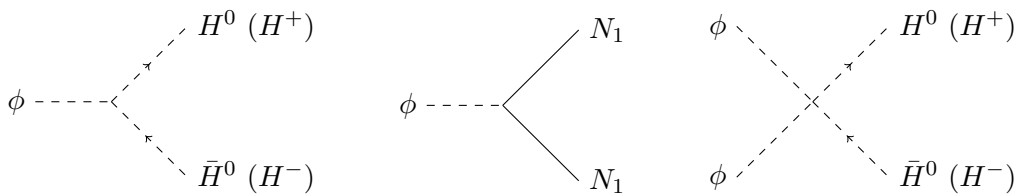

	In our scenario extending the standard leptogenesis, the additional real scalar $\phi$ could decay into RHNs. For simplicity, we assume a mass hierarchy of
	\begin{equation}
		2M_{N_1}<m_\phi<2M_{N_2}\ll M_{N_3},
	\end{equation}
	which implies that $\phi$ can only decay into $N_1 N_1$ and $H\bar{H}$.
	The relevant decay and annihilation diagrams for $\phi$ are illustrated in Fig.~\ref{feynman}.	
	At tree level, the partial decay widths for $\phi \to N_1 N_1$ and $\phi \to H \bar{H}$ are
	\begin{eqnarray}
		\Gamma(\phi\to N_1N_1)&=&\frac{y_1^2m_\phi}{16\pi}\left(1-\frac{4M_{N_1}^2}{m_\phi^2}\right)^{3/2},\\
		\Gamma(\phi\to H\bar{H})&\simeq&\frac{\kappa^2}{8\pi m_\phi}.
	\end{eqnarray}
	The annihilation cross sections for $\phi\phi \to H\bar{H}$ are given by
	\begin{equation}
		\sigma(\phi\phi\to H^0\bar{H}^0)=\sigma(\phi\phi\to H^+H^-)\simeq\frac{\lambda_{\phi H}^2}{4\pi s\sqrt{1-4m_\phi^2/s}},
	\end{equation}
	where $s$ is a Mandelstam variable.
	
	The $\phi$ decays into $N_1 N_1$ are followed by the subsequent $N_1$ decays, which give rise to the lepton asymmetry if the $N_1$ particles depart from thermal equilibrium.
	If the $\phi$ lifetime is sufficiently long, the production of $N_1$ particles persists even when the washout effect is suppressed at low temperatures.
	In the next section, we will provide a comprehensive analysis for the evolution of the lepton asymmetry.
	
\section{Boltzmann equations}
\label{sec:Boltzmann}
	
In this section, we utilize a set of Boltzmann equations~\cite{Kolb:1990vq, Xing:2011zza, Bodeker:2020ghk} to study the evolution of the number densities
of the $\phi$ particles, the $N_1$ particles, and the $B-L$ charges, which are denoted by $n_\phi$, $n_{N_1}$, and $n_{B-L}$, respectively.
With the assumptions made in the previous section, the Boltzmann equation for $n_\phi$ reads
\begin{eqnarray}
\frac{dn_\phi}{dt} + 3Hn_\phi &=& -\langle\Gamma_{\phi\to N_1N_1}\rangle\left[n_\phi-\frac{n_\phi^\mathrm{eq}n_{N_1}^2}{(n_{N_1}^\mathrm{eq})^2}\right]-\langle\Gamma_{\phi \to H \bar{H}}\rangle\left(n_\phi-n_\phi^\mathrm{eq}\right)
\nonumber\\
&& -\langle\sigma_{\phi\phi\to H \bar{H}}v\rangle\left[n_\phi^2-(n_\phi^\mathrm{eq})^2\right],
\label{eq:n_phi}
\end{eqnarray}
where $H$ is the Hubble expansion rate, and the collision terms are related to the $\phi \leftrightarrow N_1 N_1$, $\phi \leftrightarrow H \bar{H}$, and $\phi\phi \leftrightarrow H \bar{H}$ processes.
$n_i^\mathrm{eq}$ indicate the number densities in thermal equilibrium, while $\langle \Gamma_i\rangle$ and $\langle \sigma_{\phi\phi\to H \bar{H}} v\rangle$ denote the thermal averages of the decay width $\Gamma_i$ and the $\phi\phi\to H \bar{H}$ scattering cross section $\sigma_{\phi\phi\to H \bar{H}}$ multiplied by the M\o ller velocity, respectively.

Taking into account the $N_1 \leftrightarrow \ell H / \bar{\ell}\bar{H}$ and $\phi \leftrightarrow N_1 N_1$ processes, the Boltzmann equation for $n_{N_1}$ is
\begin{equation}\label{eq:n_N1}
\frac{dn_{N_1}}{dt} + 3Hn_{N_1} = -\langle\Gamma_{N_1}\rangle\left(n_{N_1}-n_{N_1}^\mathrm{eq}\right) + \langle\Gamma_{\phi\to N_1 N_1}\rangle\left[n_\phi-\frac{n_\phi^\mathrm{eq}n_{N_1}^2}{(n_{N_1}^\mathrm{eq})^2}\right].
\end{equation}
Moreover, the evolution of $n_{B-L}$ is governed by
\begin{equation}\label{eq:n_B-L}
\frac{dn_{B-L}}{dt} + 3Hn_{B-L} = -\varepsilon_1 \langle\Gamma_{N_1}\rangle\left(n_{N_1} -n_{N_1}^\mathrm{eq}\right)-\frac{1}{2}\langle\Gamma_{N_1}\rangle \frac{n_{N_1}^\mathrm{eq} n_{B-L}}{n_\ell^\mathrm{eq}},
\end{equation}
where the first term in the right-hand side corresponds to the generation of the lepton asymmetry by the $CP$ asymmetry $\varepsilon_1$ and the departure of the $N_1$ particles from thermal equilibrium, and the second term, which is proportional to $n_{B-L}$, indicates the washout of the generated lepton asymmetry caused by the inverse decays.

Note that the integrated Boltzmann equations  \eqref{eq:n_phi}--\eqref{eq:n_B-L} are derived from the full Boltzmann equations that govern the evolution of momentum distributions of particles~\cite{Basboll:2006yx,Garayoa:2009my,Hahn-Woernle:2009jyb}, assuming that the $\phi$ and $N_1$ particles are in kinetic equilibrium with the SM plasma at the same temperature.
This assumption ensures that the decay width or cross section of a process, along with the corresponding equilibrium number densities, can be used to describe the contribution of the inverse process.
For instance, the contributions of $N_1 N_1 \to \phi$ and $H \bar{H} \to \phi\phi$ in Eq.~\eqref{eq:n_phi} are given by $\langle\Gamma_{\phi\to N_1N_1}\rangle {n_\phi^\mathrm{eq}n_{N_1}^2}/{(n_{N_1}^\mathrm{eq})^2}$ and $\langle\sigma_{\phi\phi\to H \bar{H}}v\rangle (n_\phi^\mathrm{eq})^2$, respectively.
If the kinetic equilibrium is not maintained, the integrated Boltzmann equations cannot be expressed in such closed forms, and the full Boltzmann equations are more appropriate for obtaining precise results.

Once the $\phi$ particles are produced via the $H\bar{H} \to \phi\phi$ process at high temperatures $T \gg m_\phi$, they rapidly archive kinetic equilibrium through the scattering processes $\phi H \to \phi H$ and $\phi \bar{H} \to \phi \bar{H}$, primarily mediated by the quartic coupling $\lambda_{\phi H}$.
Nonetheless, for low temperatures $T \lesssim m_\phi$, the kinetic equilibrium may no longer be maintained, and its condition must be carefully examined.
The scattering cross section for $\phi H \to \phi H$ with nonrelativistic $\phi$ particles is given by
\begin{equation}
	\langle\sigma_\mathrm{scat} v\rangle \simeq  \frac{\lambda_{\phi H}^2}{16\pi m_\phi^2},
\end{equation}
which leads to a scattering rate of $\Gamma_\mathrm{scat} = 4 n_H \langle\sigma_\mathrm{scat} v\rangle$, accounting for the four degrees of freedom of the Higgs bosons.
Based on analyses in Refs.~\cite{Hofmann:2001bi,Visinelli:2015eka,Cai:2021wmu}, the condition for kinetic equilibrium can be approximated as
\begin{equation}\label{eq:KE_condition}
	\frac{T}{m_\phi}\frac{\Gamma_\mathrm{scat}}{H} \gtrsim 1,
\end{equation}
where the suppression factor $T/m_\phi$ reflects  the requirement of a sufficient number of scatterings to establish kinetic equilibrium.

For our scenario to be viable, the quartic coupling $\lambda_{\phi H}$ cannot be too large; otherwise, the annihilation process $\phi \phi \to H \bar{H}$ would excessively deplete the $\phi$ particles, rendering the nonthermal production of $N_1$ particles via $\phi$ decays inefficient.
For the values of $\lambda_{\phi H}$ used in the following analysis, the condition \eqref{eq:KE_condition} can hardly be satisfied for $T \lesssim m_\phi$.
Consequently, at such low temperatures, the kinetic equilibrium of $\phi$ particles with the SM plasma breaks down.
This implies that the contributions of the inverse processes $N_1 N_1 \to \phi$, $H \bar{H}\to \phi$, and $H \bar{H} \to \phi\phi$ no longer retain the closed forms presented in Eqs.~\eqref{eq:n_phi} and \eqref{eq:n_N1}.
Nevertheless, these contributions are greatly suppressed for $T \ll m_\phi$, as the $N_1$ particles and Higgs bosons lack sufficient energy to produce $\phi$ particles.
As a result, the integrated Boltzmann equations \eqref{eq:n_phi} and \eqref{eq:n_N1} still serve as credible approximations for $T \lesssim m_\phi$.

The kinetic equilibrium of $N_1$ particles is primarily maintained by the $1\leftrightarrow 2$ processes $N_1 \leftrightarrow \ell H / \bar{\ell}\bar{H}$ and $\phi \leftrightarrow N_1 N_1$.
In contrast to $2\leftrightarrow 2$ processes, $1\leftrightarrow 2$ processes are generally inefficient for achieving thermalization, rendering the assumption of kinetic equilibrium largely invalid~\cite{Basboll:2006yx,Garayoa:2009my,Hahn-Woernle:2009jyb}.
Nevertheless, solutions to the full Boltzmann equations indicate that the resulting lepton asymmetry does not differ significantly from that obtained by solving the integrated Boltzmann equations.
If one defines $K \equiv \Gamma_{N_1}/H(M_{N_1})$, where $H(M_{N_1})$ is the Hubble rate at $T = M_{N_1}$, the difference of the lepton asymmetry for $K > 5$ is less than $15\%$~\cite{Basboll:2006yx}.
For the parameter points used in the subsequent analysis, the condition $K > 5$ is always satisfied.
Therefore, the results and conclusions derived from solving the integrated Boltzmann equations \eqref{eq:n_phi}--\eqref{eq:n_B-L} are expected to be reliable.

In order to separate the dynamical evolution from the cosmological expansion effect, it is more convenient to use the ratios of these number density to the photon number density, $Y_i = n_i/n_\gamma$.
A dimensionless parameter $x = M_{N_1} / T$ is further introduced to replace the temperature $T$.
Thus, the set of Boltzmann equations \eqref{eq:n_phi}--\eqref{eq:n_B-L} can be transformed into a simplified form,
\begin{eqnarray}
\frac{dY_\phi}{dx} &=& -\frac{ \langle \Gamma_{\phi\to N_1 N_1}\rangle}{Hx}\left[Y_\phi-\frac{Y_\phi^\mathrm{eq}Y_{N_1}^2}{(Y_{N_1}^\mathrm{eq})^2}\right]
-\frac{ \langle \Gamma_{\phi\to H \bar{H}}\rangle}{Hx}\left(Y_\phi-Y_\phi^\mathrm{eq}\right)
\nonumber\\
&& -\frac{ n_\gamma\langle\sigma_{\phi\phi\to H\bar{H}} v\rangle}{Hx} \left[Y_\phi^2-(Y_\phi^\mathrm{eq})^2\right],\label{Y_phi}
\\ 
\frac{dY_{N_1}}{dx} &=& -\frac{ \langle \Gamma_{N_1}\rangle}{Hx}\left(Y_{N_1}-Y_{N_1}^\mathrm{eq}\right) + \frac{ \langle \Gamma_{\phi\to N_1 N_1}\rangle}{Hx}\left[Y_\phi-\frac{Y_\phi^\mathrm{eq}Y_{N_1}^2}{(Y_{N_1}^\mathrm{eq})^2}\right],
\\
\frac{dY_{B-L}}{dx} &=& -\frac{ \varepsilon_1\langle \Gamma_{N_1}\rangle}{Hx} \left(Y_{N_1}-Y_{N_1}^\mathrm{eq}\right)-\frac{ \langle\Gamma_{N_1}\rangle Y_{N_1}^\mathrm{eq} }{2Hx Y_\ell^\mathrm{eq}}\,Y_{B-L} ,
\label{Y_B-L}
\end{eqnarray}
where $Y_i^\mathrm{eq} \equiv n_i^\mathrm{eq}/n_\gamma$.
Based on the Friedmann equation, the Hubble rate $H$ is determined by the energy densities of SM radiation, $\phi$, and $N_1$:
\begin{eqnarray}
    H^2 = \frac{8\pi}{3M_\mathrm{Pl}^2}(\rho_\mathrm{R} + \rho_\phi + \rho_{N_1}).
\end{eqnarray}
The SM radiation energy density is
\begin{eqnarray}
    \rho_\mathrm{R} = \frac{\pi^2}{30}\, g_* T^4,
\end{eqnarray}
where $g_*$ is the SM effective relativistic degrees of freedom.

Below we solve the set of Boltzmann equations by numerical methods.
We assume that after reheating, all SM particles are in thermal equilibrium, while there are no $N_i$ and $\phi$ particles because their interactions are too weak.
Thus, the initial conditions can be set as 
$Y_\phi =Y_{N_1}=Y_{B-L}=0$ at $x = 0.01$.
Subsequently, $\phi$ particles are produced through interactions with SM Higgs bosons.
For the interested parameter regions discussed below, they are nonrelativistic at relevant temperatures and hardly reach thermal equilibrium, so their energy density can be approximated by $\rho_\phi \simeq m_\phi n_\phi$.

Furthermore, $\langle \Gamma_i\rangle$ in the Boltzmann equations can be expressed as~\cite{Kolb:1979qa, Buchmuller:2004nz}
\begin{equation}
\langle \Gamma_i\rangle = \frac{\mathrm{K}_1(x)}{\mathrm{K}_2(x)}\,\Gamma_i,
\end{equation}
where $\mathrm{K}_j(x)$ represents the the second kind modified Bessel function of order $j$.
$\langle\sigma_{\phi\phi\to H \bar{H}} v\rangle$ can be calculated by~\cite{Gondolo:1990dk, Plumacher:1997ru}
\begin{equation}
\langle\sigma_{\phi\phi\to H \bar{H}} v\rangle = \frac{T}{32\pi^4 (n^\mathrm{eq}_\phi)^2}\int^\infty_{4m_\phi^2} ds\, \sqrt{s} (s-4m_\phi^2) \,\sigma_{\phi\phi\to H \bar{H}}(s)\, \mathrm{K}_1\left(\frac{\sqrt{s}}{T}\right).
\end{equation}
The washout effect in Eq.~\eqref{Y_B-L} is related to a factor of~\cite{Buchmuller:2004nz}
\begin{equation}
W_\mathrm{ID}(x) = \frac{\langle\Gamma_{N_1}\rangle}{2Hx}\frac{Y_{N_1}^\mathrm{eq}}{Y_\ell^\mathrm{eq}}=\frac{x \mathrm{K}_1(x)\Gamma_N }{4 H}.
\end{equation} 	
 
When the $\phi$ decay processes are not effective, the energy density of nonrelativistic $\phi$ particles scale with the scale factor $a$ as $\rho_\phi \propto a^{-3}$ because of the cosmic expansion.
Compared with the radiation energy density $\rho_\mathrm{R} \propto a^{-4}$, the dilution of $\phi$ particles are slower.
Consequently, if the $\phi$ decay width is too small to deplete $\phi$ particles early enough, $\rho_\phi$ would soon dominate the universe, leading to an early matter-domination era that makes the evolution of the universe deviated from the standard $\Lambda\mathrm{CDM}$ model.
This era ends after the $\phi$ decays become significant, i.e., $\langle\Gamma_\phi\rangle > H$, and additional entropy is injected by $\phi$ decays, heating up the plasma and increasing the photon number density $n_\gamma$.
On the other hand, the generation of the lepton asymmetry arising from $\phi$ decays primarily depends on the $\phi$ number density rather than on the decay width $\Gamma_\phi$.
Consequently, the resulting $Y_{B-L} = n_{B-L}/n_\gamma$ decreases, leading to a reduction in the final baryon asymmetry $\eta_B$~\cite{Cataldi:2024bcs}.
Such a dilution of the leptogenesis yield caused by entropy injection contradicts our goal of enhancing lepton asymmetry generation.
Thus, we must prevent this situation by ensuring that the entropy injection from $\phi$ decays remains negligible.
The ratio $\rho_\phi/\rho_\mathrm{R}$ reaches the maximum roughly at the time with $\langle\Gamma_\phi\rangle = H$, after which it declines due to $\phi$ decays.
Therefore, if we require $\rho_\phi/\rho_\mathrm{R} < 0.1$ when $\langle\Gamma_\phi\rangle = H$, $\phi$ particles would never constitute a significant fraction of the total energy density of the universe, and the entropy injection into the plasma due to $\phi$ decays would be negligible.

\begin{figure}[!t]
\centering
\subfigure{\includegraphics[width=0.48\textwidth]{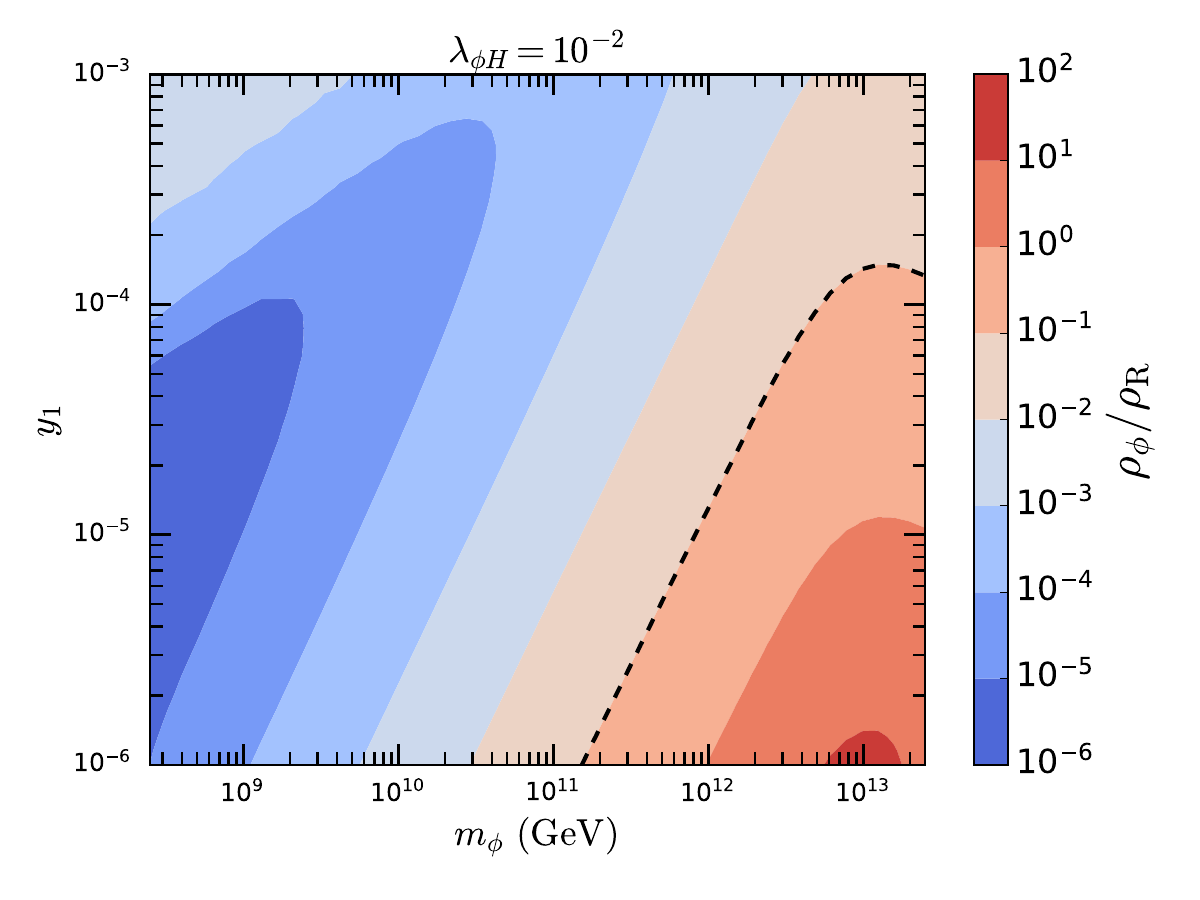}}
		\hspace{.01\textwidth}
\subfigure{\includegraphics[width=0.48\textwidth]{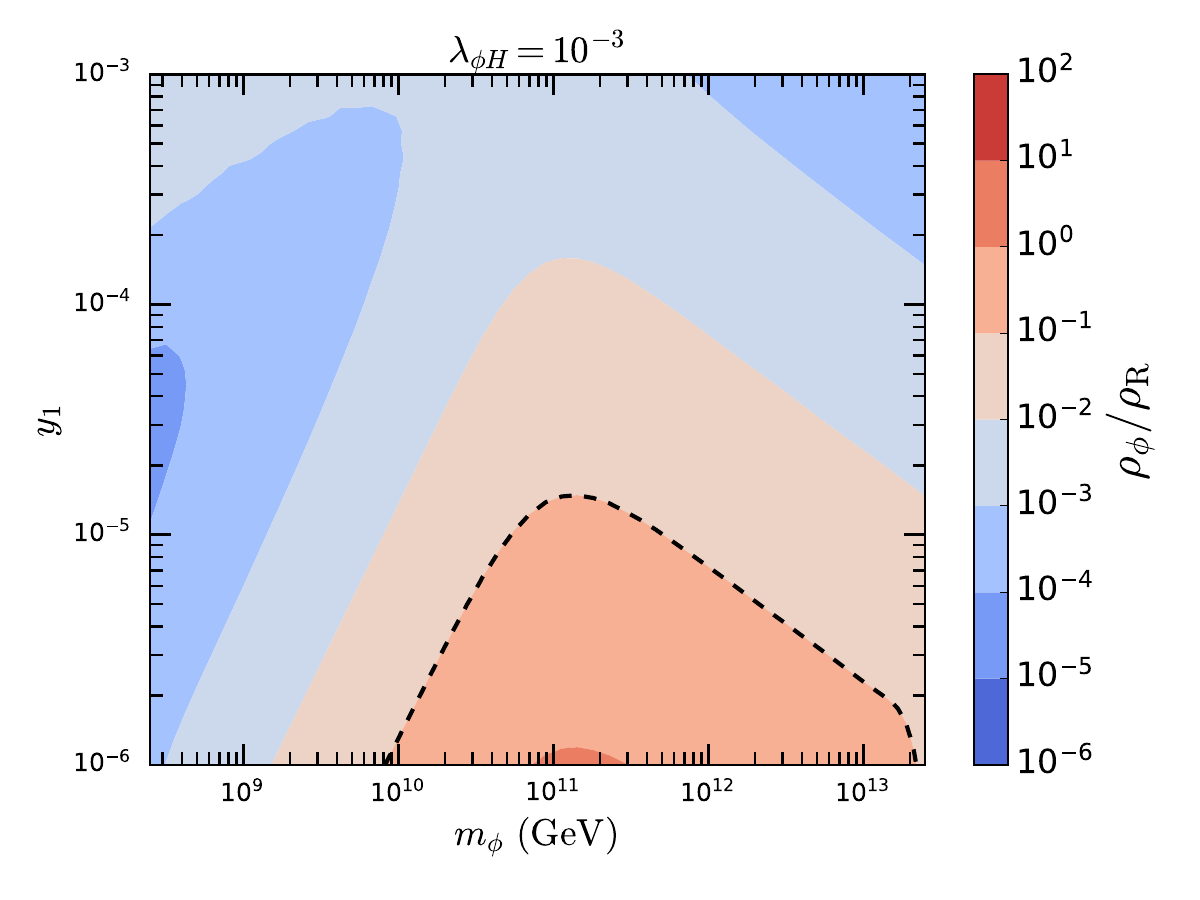}}
\caption{Contours of $\rho_\phi/\rho_R$ when $\langle\Gamma_\phi\rangle = H$ for $\lambda_{\phi H} = 10^{-2}$ (left) and $10^{-3}$ (right). The other parameters are set to be $z = 0.2+0.2i$, $\kappa = 100\mathrm{GeV}$, and $M_{N_1} = m_\phi/2.5$.
The black dashed lines represent $\rho_\phi/\rho_\mathrm{R}= 0.1$.}
\label{gammaphi}
\end{figure}

In Fig.~\ref{gammaphi}, we demonstrate the contours of $\rho_\phi/\rho_\mathrm{R}$ when $\langle\Gamma_\phi\rangle = H$ in the $m_\phi$-$y_1$ plane for $\lambda_{\phi H} = 10^{-2}$ and $10^{-3}$, and the rest of parameters are fixed as $z = 0.2+0.2i$, $\kappa = 100~\mathrm{GeV}$, and $M_{N_1} = m_\phi/2.5$.
In the parameter regions with $\rho_\phi/\rho_\mathrm{R} > 1$, there would be an early matter-domination era.
For the $\lambda_{\phi H} = 10^{-2}$ case, this occurs at a region with $y_1 \lesssim 10^{-5}$ and $m_\phi \gtrsim 10^{12}~\si{GeV}$.
For the $\lambda_{\phi H} = 10^{-3}$ case, the production rate of $\phi$ particles is lower, and $\phi$ particles can dominate the universe only in a small parameter region if $y_1 > 10^{-6}$.
In the parameter regions with $\rho_\phi/\rho_\mathrm{R} < 0.1$, the entropy injection from $\phi$ decays can be neglected, and this is the only situation we focus on in the following analysis.

We choose two benchmark points (BPs) for the model parameters, as outlined in the Table~\ref{tab:BPs}.
The Yukawa coupling matrix $h_\nu$ is determined by the Casas-Ibarra parametrization.
We present the values of the $h_\nu$ elements, the $CP$ asymmetry $\varepsilon_1$, and the $N_1$ decay width $\Gamma_{N_1}$ calculated for the two BPs in Table~\ref{tab:h_epsilon}.
Compared to BP~I with $z_\mathrm{i} = 3$, BP~II with $z_\mathrm{i} = 0.3$ has a smaller $CP$ asymmetry $\varepsilon_1$ and a much smaller $\Gamma_{N_1}$.
Thus, the washout effect in BP~II is expected to be much weaker than in BP~I.

\begin{table}[!t]
\centering
\setlength\tabcolsep{.5em}
\renewcommand{\arraystretch}{1.3}
\caption{Information for two BPs.}\label{tab:BPs}  
\begin{tabular}{ccccccccc}
\hline\hline
 & $M_{N_1}$ (GeV) & $M_{N_2}$ &$z_\mathrm{r}$&$z_\mathrm{i}$& $m_\phi$ &$y_1$ & $ \kappa$ (GeV) & $\lambda_{\phi H}$ \\
\hline
BP I	& $10^{10}$ & $10 M_{N_1}$ & 0.2 & 3 & $2.5M_{N_1}$ & $3.22\times 10^{-5}$& 100 & $6.1\times 10^{-4}$ \\
BP II	& $5\times10^9$ & $10 M_{N_1}$ & 0.45 & 0.2 & $2.5M_{N_1}$ & $3\times 10^{-5}$& 100 & $4.2\times 10^{-4}$ \\
\hline\hline
\end{tabular}
\end{table}

\begin{table}[!t]
\centering
\setlength\tabcolsep{.6em}
\renewcommand{\arraystretch}{1.3}
\caption{Values of $h_\nu$, $\varepsilon_1$, and $\Gamma_{N_1}$ for two BPs.}
\label{tab:h_epsilon}  
\begin{tabular}{cccc}
\hline\hline
& $h_\nu$ & $\varepsilon_1$&$\Gamma_{N_1}$ (GeV)\\
\hline\noalign{\medskip}
BP I & $\begin{pmatrix}
				0.42+0.54i & -1.72+1.31i & 0.01\\
				1.60-2.84i &9.03+5.04i & 0\\
				-0.45-2.82i&8.97-1.41i &0
			\end{pmatrix} \times 10^{-2}$  
&$ 3.17\times 10^{-7}$ &\ $7.67\times 10^5$\\[2.5em]
			BP II &
			$\begin{pmatrix}
				-0.01+0.40i & 0.54-1.75i &0.05i\\
				0.37+1.60i & -1.07+5.26i &-0.02i\\
				0.44+0.18i&-0.17+6.39i & 0.03i
			\end{pmatrix} \times 10^{-3}$  
			&$ 2.03\times 10^{-7}$&\ 616.89 \\
\noalign{\medskip}
\hline\hline
\end{tabular}
\end{table}

\begin{figure}[!t]
\centering
\subfigure{\includegraphics[width=0.48\textwidth]{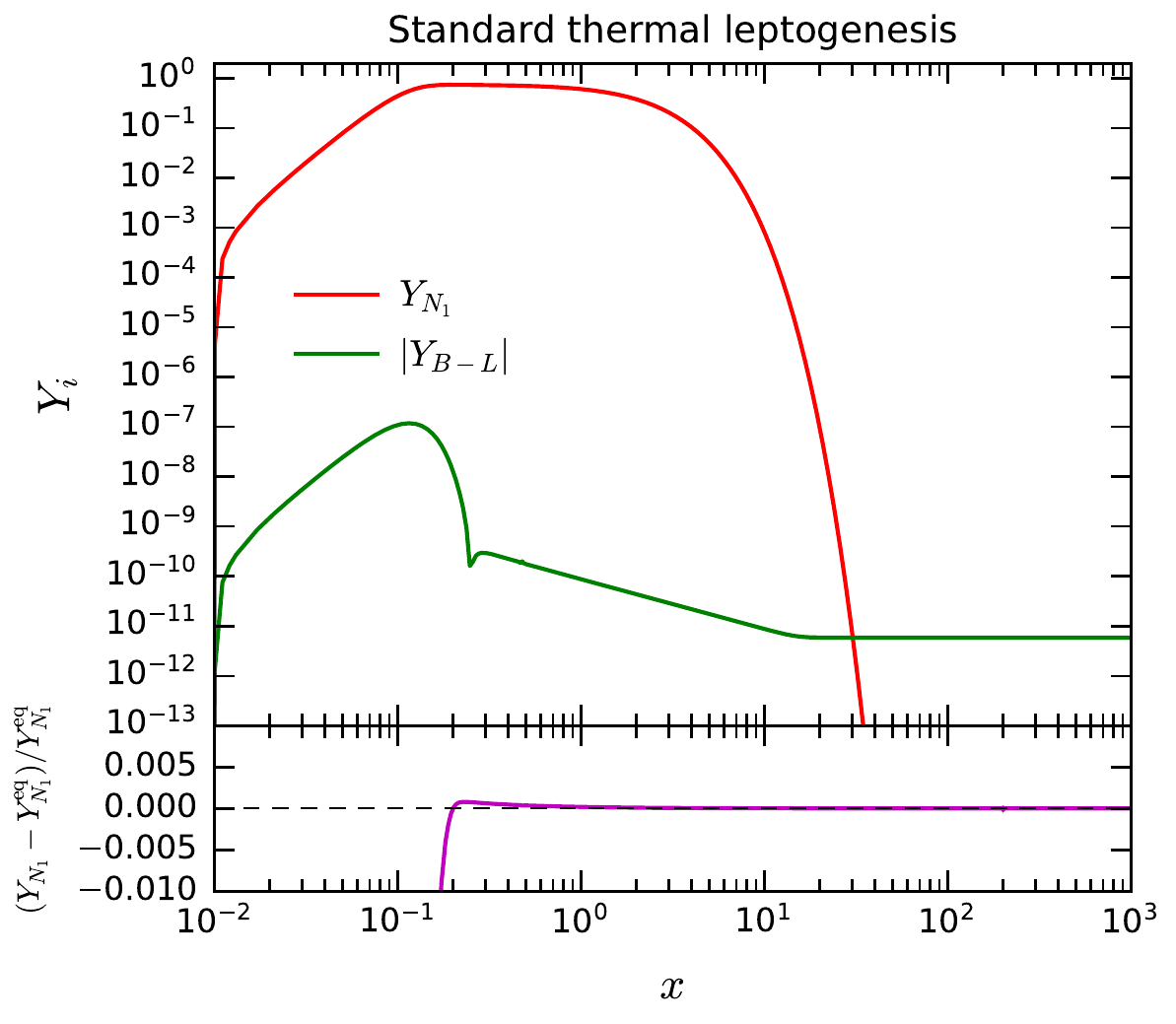}}
		\hspace{.01\textwidth}
\subfigure{\includegraphics[width=0.48\textwidth]{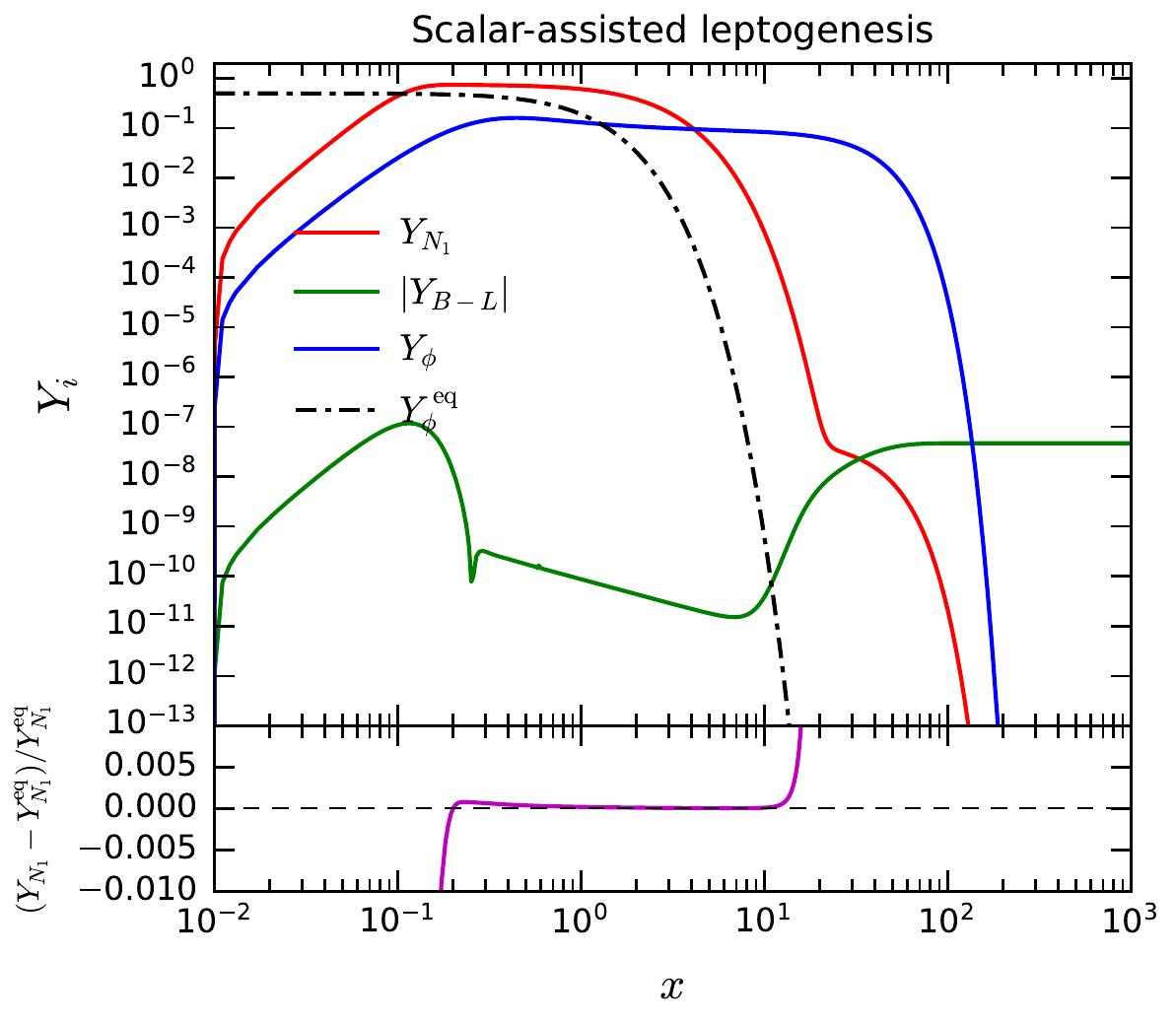}}
\caption{Evolution of $Y_{N_1}$, $Y_{B-L}$, $(Y_{N_1} - Y_{N_1}^\mathrm{eq})/Y_{N_1}^\mathrm{eq}$, $Y_\phi$, and $Y_\phi^\mathrm{eq}$ as functions of $x = M_{N_1}/T$ in the standard (left) and scalar-assisted (right) leptogenesis for BP I, which is a strong washout scheme.
The final baryon-to-photon ratios are $\eta_B =7.36\times 10^{-14}$(left) and $\eta_B = 6.1\times 10^{-10}$ (right).}
\label{strong}
\end{figure}

In order to evaluate the impact of the scalar $\phi$, we separately demonstrate the evolution of number densities in the standard thermal leptogenesis without $\phi$ and that in the scalar-assisted leptogenesis.
For the strong washout scheme BP~I, $Y_{N_1}$, $|Y_{B-L}|$, and $(Y_{N_1} - Y_{N_1}^\mathrm{eq})/Y_{N_1}^\mathrm{eq}$ as functions of $x$ in standard case are illustrated in left panel of Fig.~\ref{strong}, while $Y_\phi$ and $Y_\phi^\mathrm{eq}$ are also plotted in the right panel of Fig.~\ref{strong} for the scalar-assisted case.

In the standard case of BP~I,
although we have set a vanishing initial number density for $N_1$, the $N_1$ particles are rapidly thermalized via interactions with the SM plasma, and $Y_{N_1}$ quickly approaches the thermal value $Y_{N_1}^\mathrm{eq}$.
The $CP$-violating $N_1 \leftrightarrow \ell H/\bar{\ell}\bar{H}$ processes for nonzero $Y_{N_1} - Y_{N_1}^\mathrm{eq}$ generate the lepton asymmetry, leading to the increase of $|Y_{B-L}|$ from $x = 0.01$.
Then the washout processes begin to reduce the produced lepton asymmetry, resulting in the decrease of $|Y_{B-L}|$.
There is a sharp change in the $|Y_{B-L}|$ curve at $x \sim 0.25$, and it corresponds to the sign flip of $Y_{B-L}$ because the first negative $Y_{N_1} - Y_{N_1}^\mathrm{eq}$ become positive after $x \sim 0.2$.
For $x \gtrsim 0.25$, the net lepton numbers are regenerated under the competition between the generation and washout processes.
The $Y_{B-L}$ value tends to be steady for $x \gtrsim 20$, where both the generation and washout of the lepton symmetry are suppressed at such low temperatures.
This leads to a final baryon-to-photon ratio of $\eta_B = \num{7.36e-14}$, which is too low to account for the observed value \eqref{eq:eta_B_obs}.

In the scalar-assisted case of BP~I, the $\phi$ particles never reach thermal equilibrium. They are primarily produced at high temperatures through interactions with Higgs bosons, and finally dissipated by decay processes, among which the $\phi \to N_1 N_1$ decay leads to nonthermal production of $N_1$ particles. At low temperatures, such an extra source for the $N_1$ particles helps the generation of the lepton asymmetry. As demonstrated in the right panel of Fig.~\ref{strong}, $|Y_{B-L}|$ rises again at $x \sim 8$, and becomes constant for $x \gtrsim 60$, resulting in a final $\eta_B$ of $\num{6.1e-10}$ to explain the observation.
	
\begin{figure}[!t]
\centering
\subfigure{\includegraphics[width=0.48\textwidth]{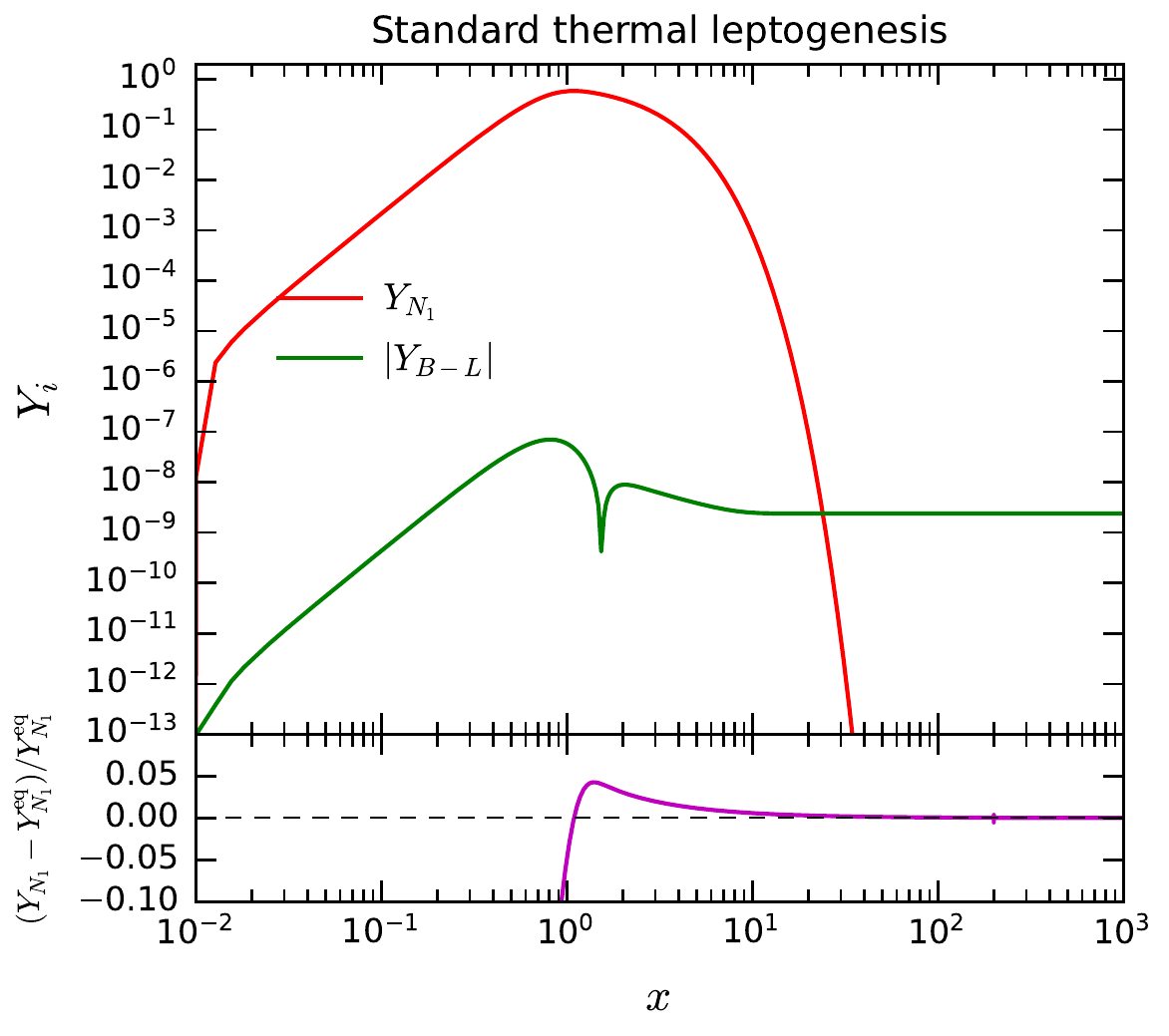}}
\hspace{.01\textwidth}
\subfigure{\includegraphics[width=0.48\textwidth]{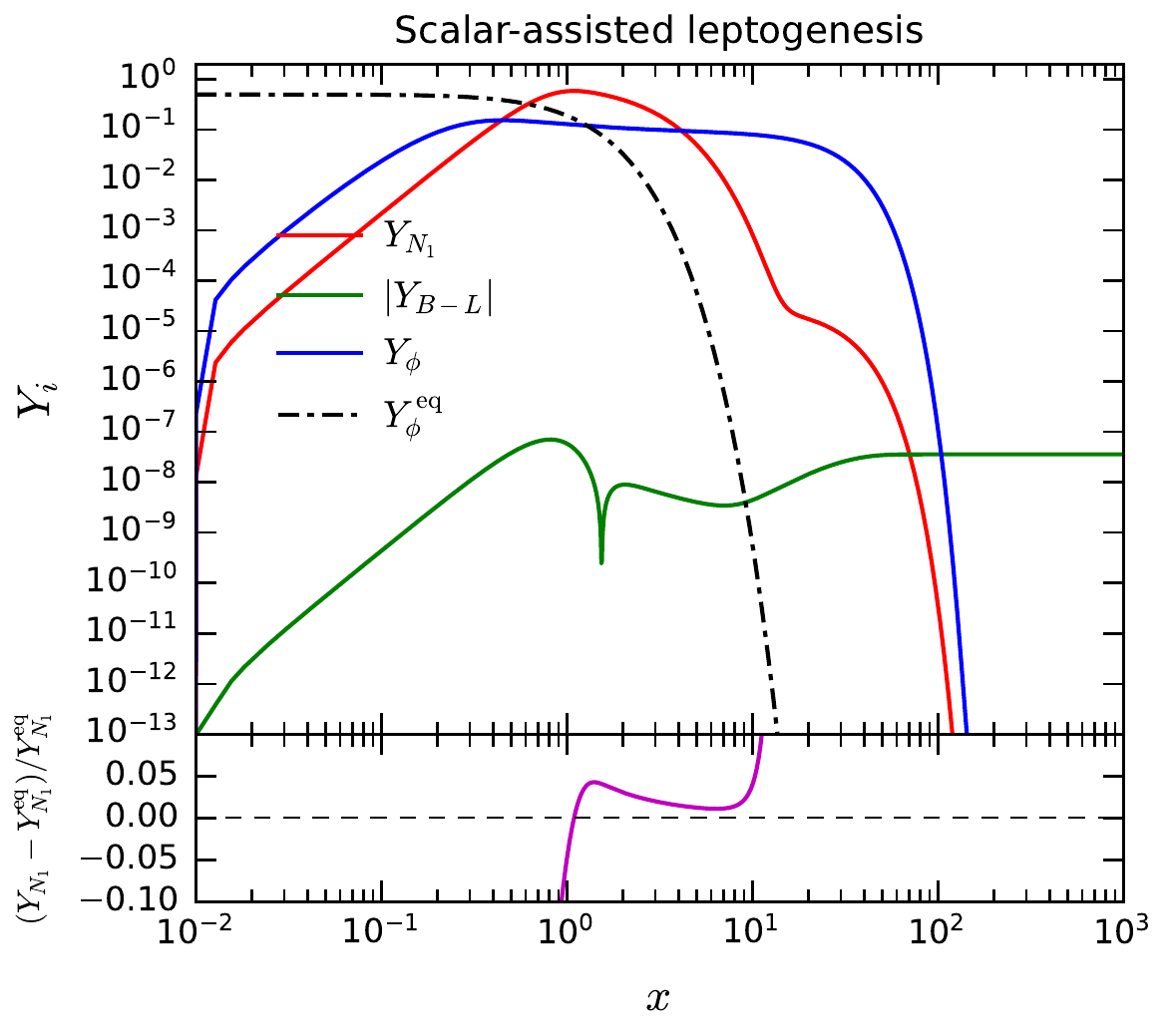}}
\caption{
Same as Fig.~\ref{strong}, but for BP~II, which is a weak washout scheme.
The final $\eta_B$ are $3.12\times 10^{-11}$ (left) and $6.1\times 10^{-10}$ (right).}
\label{weak}
\end{figure}

Similarly, we present the results for the weak washout scheme BP~II in Fig.~\ref{weak}.
Compared with BP~I, the small $\Gamma_{N_1}$ in BP~II slows down the thermalization of $N_1$ particles, and $Y_{N_1}$ becomes close to $Y_{N_1}^\mathrm{eq}$ at $x\sim 1$.
A small $\Gamma_{N_1}$ also implies a weak washout effect, and hence the decrease of $|Y_{B-L}|$ after $x \sim 2$ is quite slow.
In the standard case, the final $\eta_B$ is $\num{3.12e-11}$, lower than the observed value by one order of magnitude.
In the scalar-assisted case, $|Y_{B-L}|$ is lifted after $x \sim 10$, leading to a final $\eta_B$ of $\num{6.1e-10}$.

The discussions above show that the introduction of the scalar $\phi$ can effectively enhance the generation of the lepton asymmetry, especially in a strong washout scheme.
Such an enhancement depends on the $\phi \to N_1 N_1$ decay width, which is proportional to the square of the Yukawa coupling $y_1$.
Meanwhile, the $\phi$ number density at the time of decay impacts on the outcome, and it means that the $\phi$ coupling to the Higgs field, $\lambda_{\phi H}$, also plays an important role. Below, we analyze the individual effects of $y_1$ and $\lambda_{\phi H}$ on the final baryon-to-photon ratio $\eta_B$.

\begin{figure}[!t]
\centering
\subfigure{\includegraphics[width=0.48\textwidth]{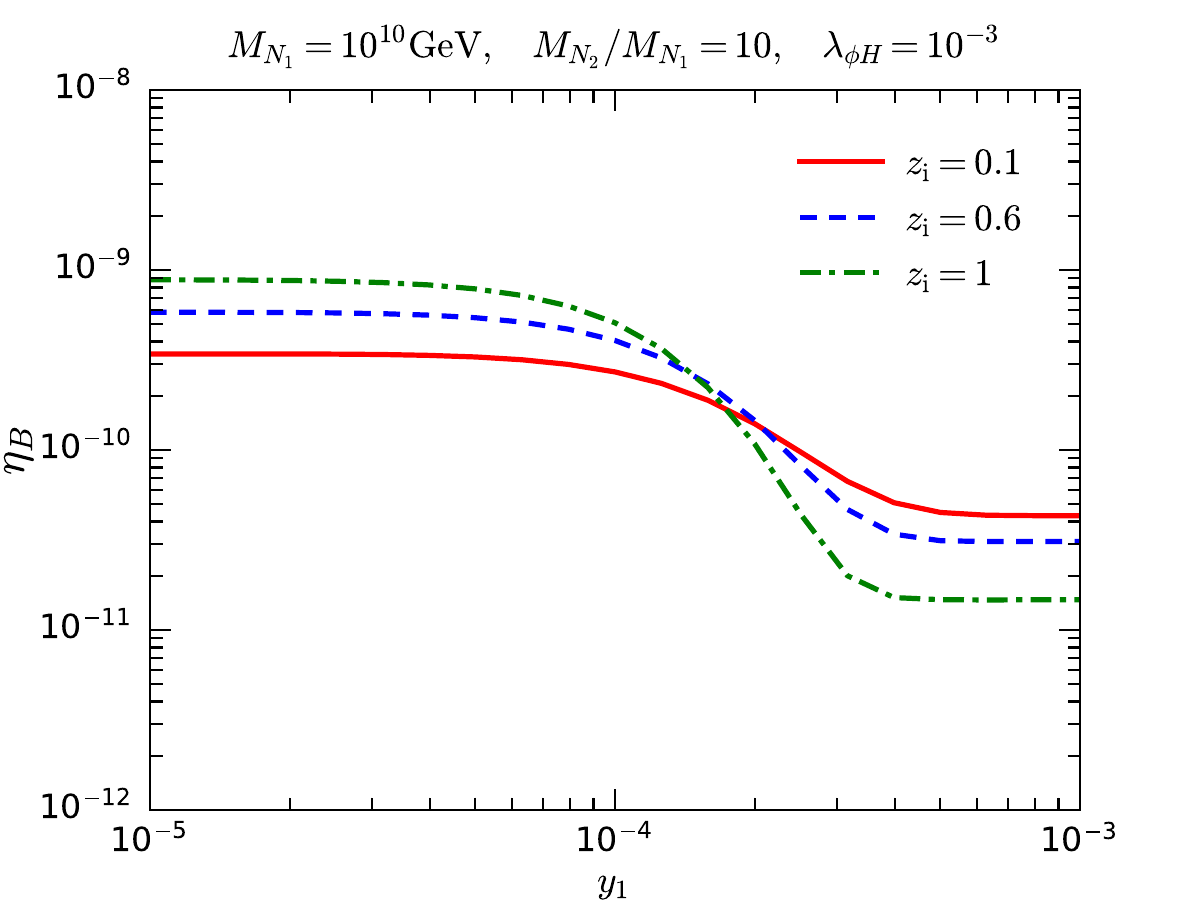}}
\hspace{.01\textwidth}
\subfigure{\includegraphics[width=0.48\textwidth]{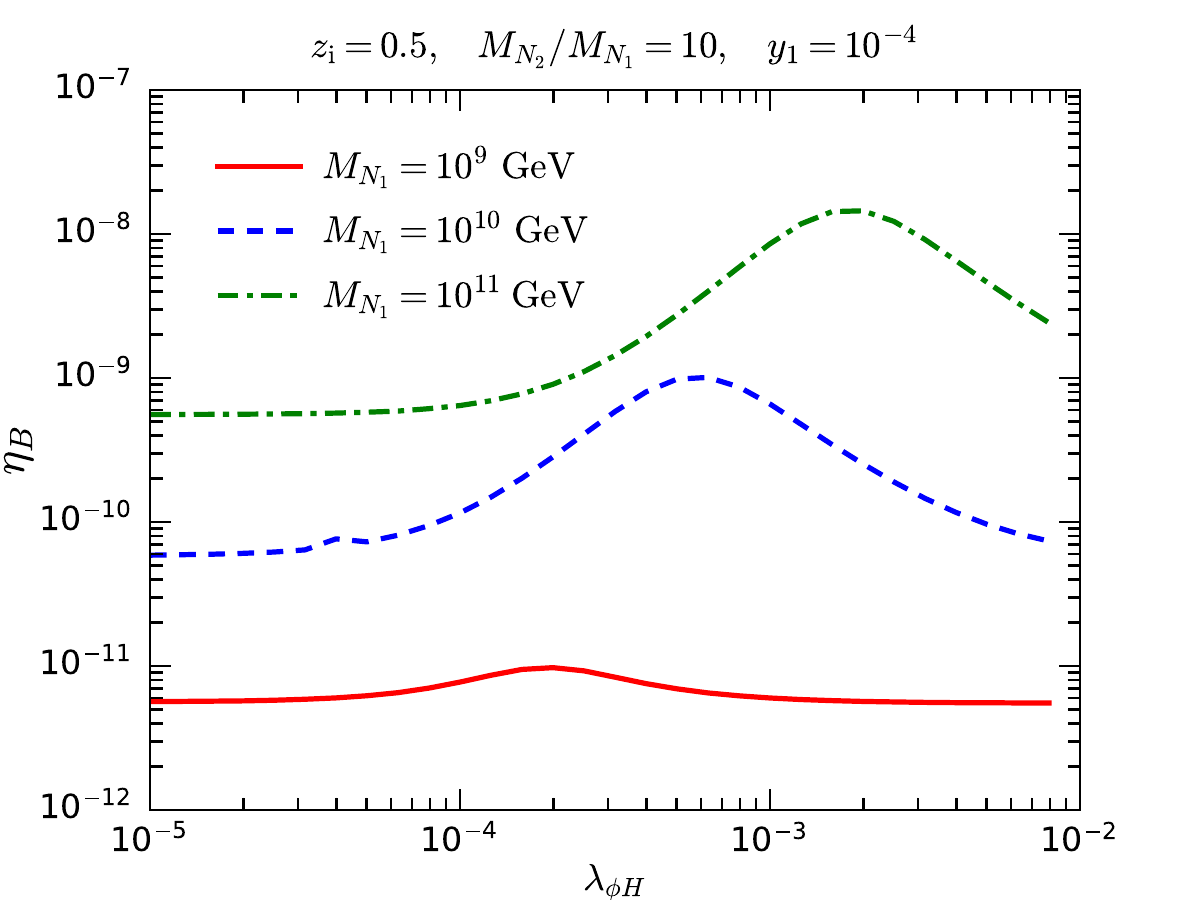}}
\caption{The baryon-to-photon ratio $\eta_B$ as functions of the Yukawa coupling $y_1$ for different values of $z_\mathrm{i}$ with $M_{N_1} = 10^{10}~\si{GeV}$ and as functions of the quartic coupling $\lambda_{\phi H}$ for different values of $M_{N_1}$ with $z_\mathrm{i} = 0.5$ and $y_1 = 10^{-4}$ (right) in the scalar-assisted leptogenesis.
The remaining parameters take the same values in BP~II.}
\label{y1eta}
\end{figure}

In the left panel of Fig.~\ref{y1eta}, we show $\eta_B$ as functions of $y_1$ for $z_\mathrm{i} = 0.1$, $0.6$, and $1$ with $M_{N_1} = 10^{10}~\si{GeV}$ in the scalar-assisted leptogenesis.
The values of the remaining parameters are the same as in BP~II, including the mass relations $M_{N_2} = 10 M_{N_1}$ and $m_\phi = 2.5 M_{N_1}$.
For a fixed $z_\mathrm{i}$, it is observed that a smaller $y_1$ leads to a larger $\eta_B$, because $\phi$ particles with a longer lifetime would maintain a high number density for a longer time to produce more $N_1$ particles.
For $y_1 \gtrsim \num{5e-4}$, the lifetime of the $\phi$ boson is too short to produce extra $N_1$ particles, and the resulting $\eta_B$ are basically the same as in the standard leptogenesis.
Note that a larger $z_\mathrm{i}$ means a larger washout effect, and the $z_\mathrm{i} = 1$ case would lead to the lowest $\eta_B$ among the three cases for $y_1 \gtrsim \num{5e-4}$.
Nonetheless, for $y_1 \lesssim \num{2e-4}$, the $\phi$ decays could overcome the strong washout in the $z_\mathrm{i} = 1$ case, making $\eta_B$ comparable to or even larger than those in the other cases.

In the right panel of Fig.~\ref{y1eta}, we similarly demonstrate $\eta_B$ as functions of $\lambda_{\phi H}$ for $M_{N_1} = 10^{9}$, $10^{10}$, and $10^{11}~\si{GeV}$ with $z_\mathrm{i} = 0.1$, $y_1 = 10^{-4}$, and other parameters taking values in BP~II.
The choice of $y_1 = 10^{-4}$ ensures $\rho_\phi/\rho_\mathrm{R} < 0.1$ when $\langle\Gamma_\phi\rangle = H$, maintaining the validity of our calculations.
It clearly shows that $\lambda_{\phi H}$ also has a significant impact on the resulting lepton asymmetry, because the $\phi$ particles are produced through interactions with the SM Higgs bosons. If the $\phi \phi \to H\bar{H}$ cross section is too high, the number density of the remaining $\phi$ particles after $\phi\phi$ annihilation would be reduced. Conversely, if the cross section is too low, fewer $\phi$ particles will be generated at the beginning. Therefore, the strong enhancement of leptogenesis by the scalar decays appears at moderate values of $\lambda_{\phi H}$, such as $\lambda_{\phi H} \sim \num{2e-3}$ for $M_{N_1} = 10^{11}~\si{GeV}$ and $\lambda_{\phi H} \sim \num{6e-4}$ for $M_{N_1} = 10^{10}~\si{GeV}$.
Since we have fixed the relation $m_\phi = 2.5 M_{N_1}$, for the case of $M_{N_1} = 10^{9}~\si{GeV}$, $\phi$ particles decay too early to make a notable contribution to $\eta_B$.

\section{Parameter scans}
\label{sec:scan}

\begin{figure}[!t]
\centering
\subfigure{\includegraphics[width=0.48\textwidth]{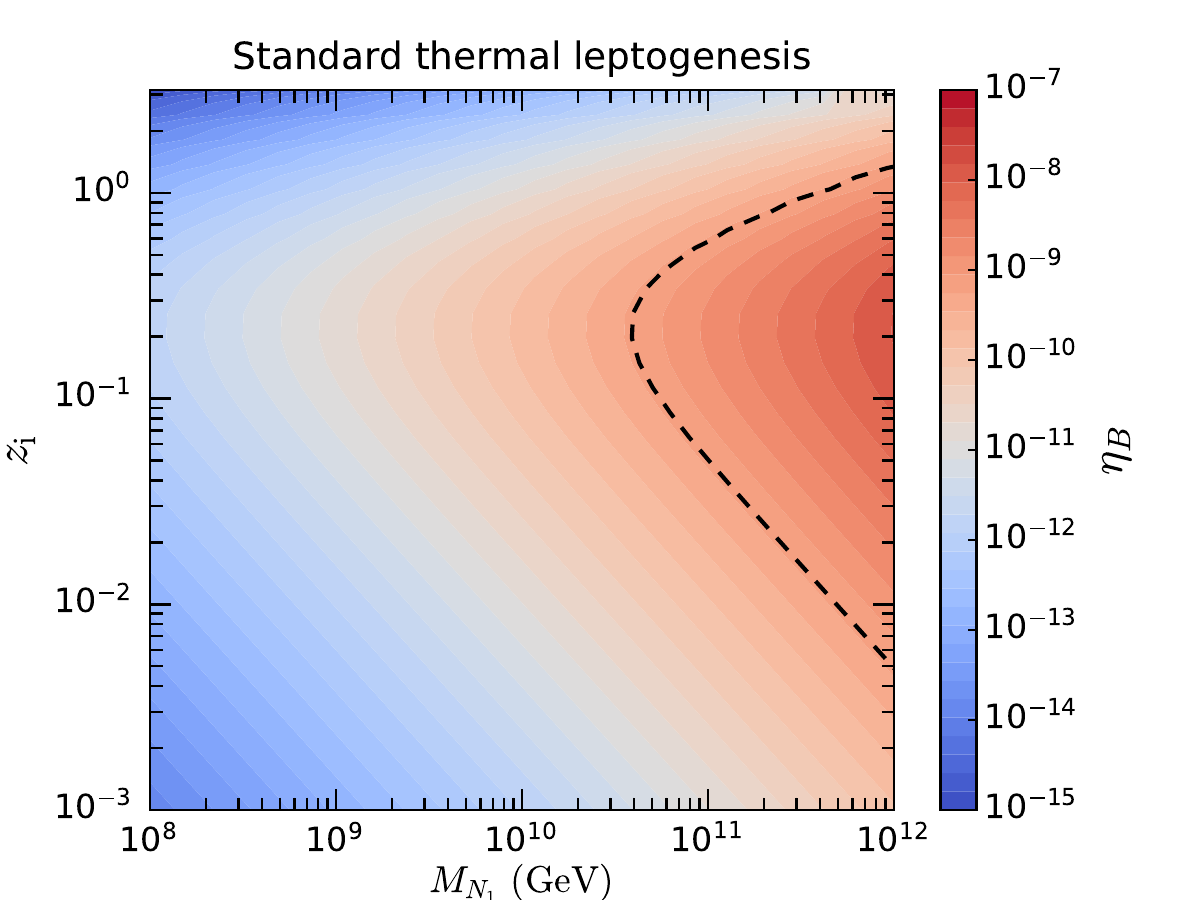}}
\hspace{.01\textwidth}
\subfigure{\includegraphics[width=0.48\textwidth]{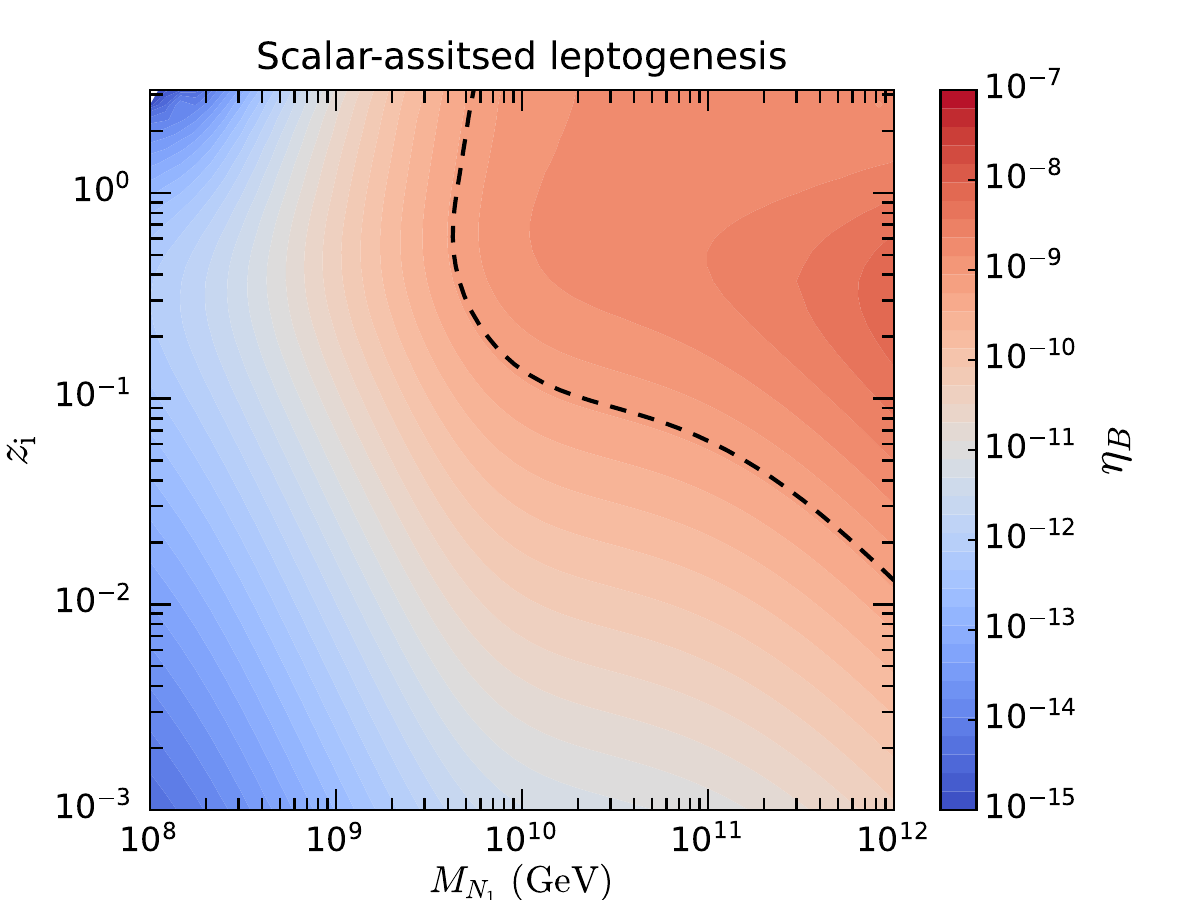}}
\caption{$\eta_B$ contours in the $M_{N_1}$-$z_\mathrm{i}$ plane for the standard (left) and scalar-assisted (right) leptogenesis with the remaining parameters taking the values in BP~II.
The dashed lines denote the observed value $\eta_B = 6.13\times10^{-10}$.}
\label{scan}
\end{figure}

In order to understand the impact of the scalar $\phi$, we perform some parameter scans in this section.
Firstly, we conduct a scan in the $M_{N_1}$-$z_\mathrm{i}$ plane with $\lambda_{\phi H}= 10^{-4}$ and $y_1=10^{-4}$, while the rest of parameters take the values in BP~II.
Contours of $\eta_B$ obtained in the standard and scalar-assisted leptogenesis are illustrated in the left and right panels of Fig.~\ref{scan}, respectively.
Note that all the parameters points in Fig.~\ref{scan} lead to $\rho_\phi/\rho_\mathrm{R} < 0.1$ when $\langle\Gamma_\phi\rangle = H$.
	
In both cases, an increase in $M_{N_1}$ implies an augmentation in the $CP$ asymmetry $\varepsilon_1$ that essentially increases $\eta_B$.
On the other hand, a larger $z_\mathrm{i}$ gives not only a larger $\varepsilon_1$, but also a larger $\Gamma_{N_1}$, which would enhance the washout effect.
Consequently, for a fixed $M_{N_1}$ in the standard case, the resulting baryon-to-photon ratio increases as $z_\mathrm{i}$ increases from $10^{-3}$, and reaches the maximum at $z_\mathrm{i} \sim 0.2$, and then declines for larger $z_\mathrm{i}$.
In the scalar-assisted case, the nonthermal production of $N_1$ particles from $\phi$ decays can significantly increase the resulting $\eta_B$.
In particular, it compensates the lepton asymmetry diminished by the strong washout for $z_\mathrm{i} \gtrsim 0.2$.
At $z_\mathrm{i} \sim 1$, the value of $M_{N_1}$ required for the correct $\eta_B$ can be reduced by nearly two orders of magnitude, compared to the standard case.

\begin{figure}[!t]
\centering
\subfigure{\includegraphics[width=0.49\textwidth]{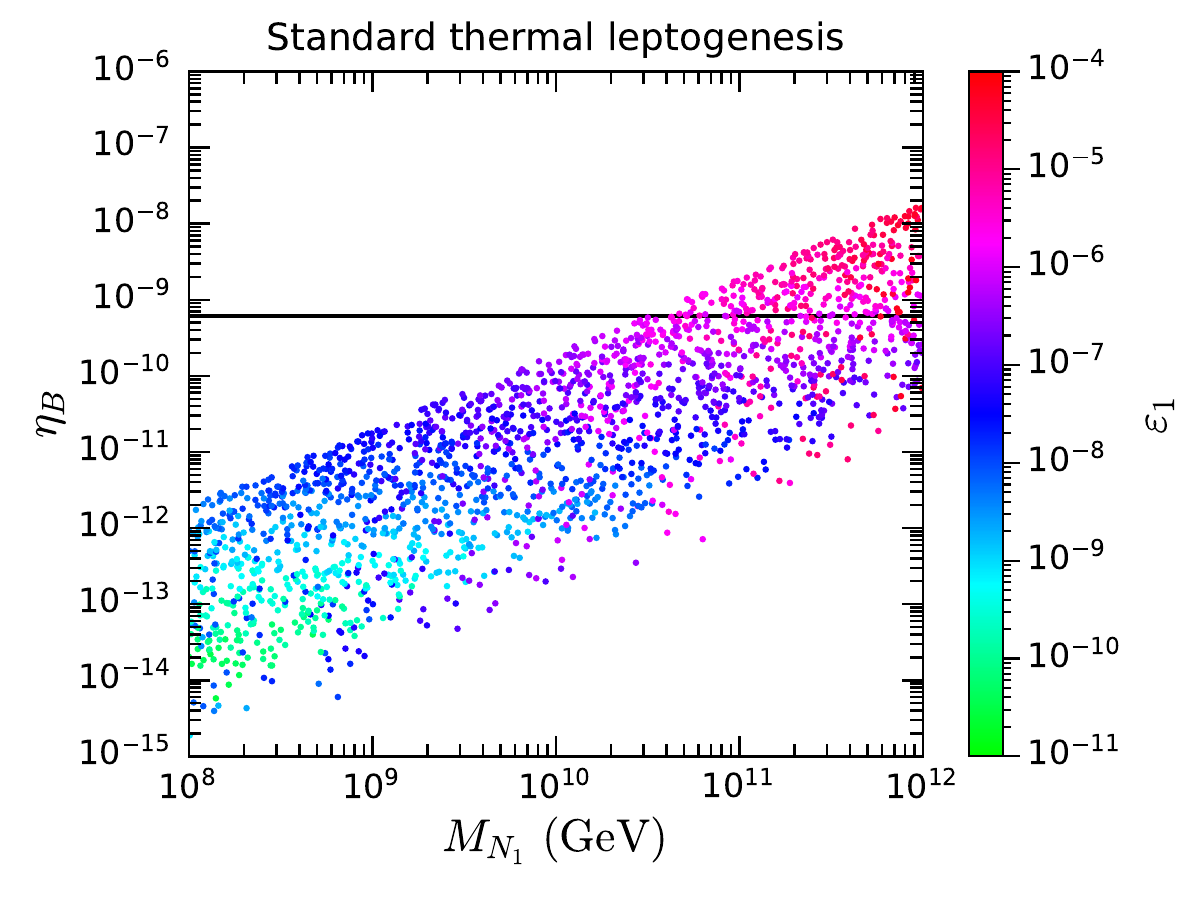}}
\subfigure{\includegraphics[width=0.49\textwidth]{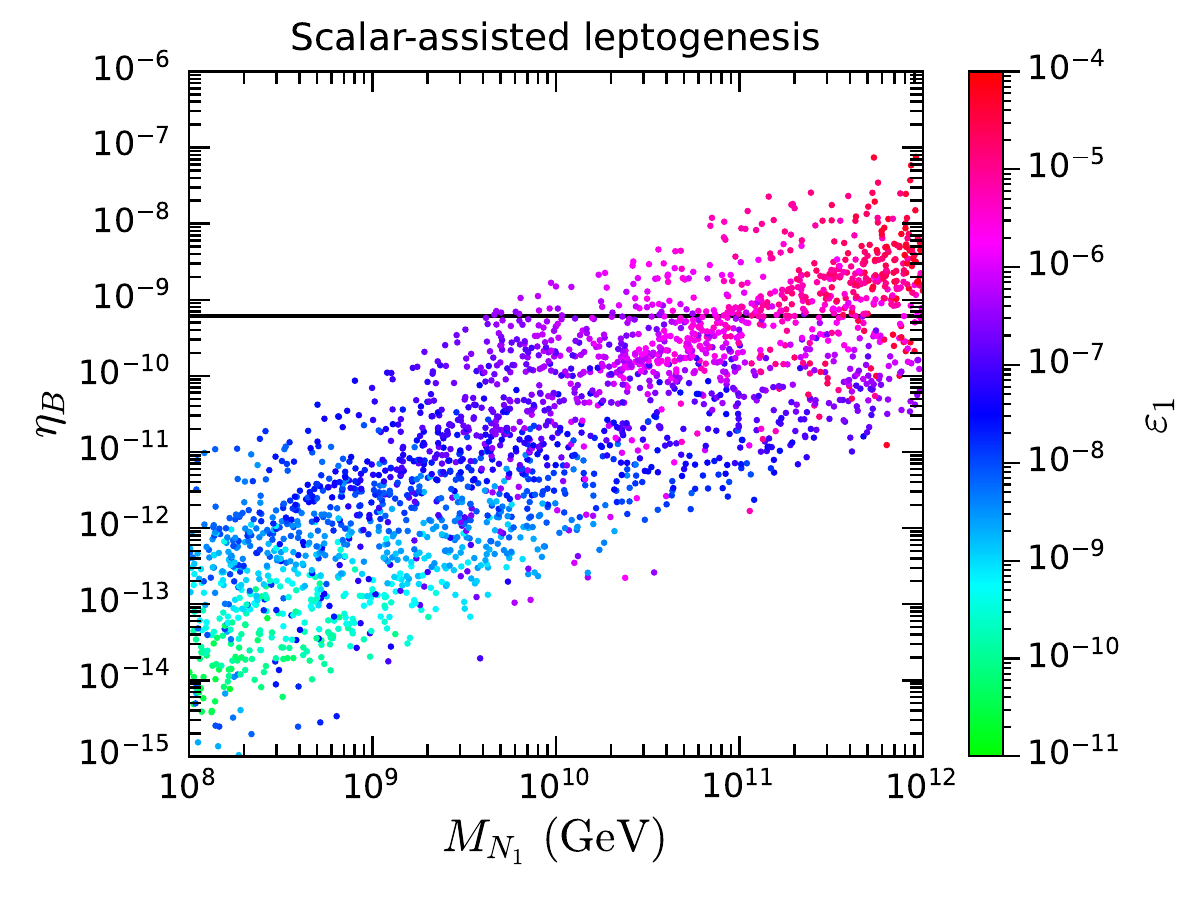}}
\caption{Parameter points projected into the $M_{N_1}$-$\eta_B$ plane for the standard (left) and scalar-assisted (right) leptogenesis.
The color axes denote the $CP$ asymmetry $\varepsilon_1$, and the black horizontal lines indicate the observed value of the baryon-to-photon ratio.}
\label{scatter0}
\end{figure}

Furthermore, we carry out a random scan within the following parameter ranges:
\begin{eqnarray}
&& 10^8~\si{GeV} < M_{N_1} < 10^{10}~\si{GeV}\quad \text{with}\quad 10^{-5}<y_1<10^{-3},
\\
&& 10^{10}~\si{GeV} < M_{N_1} < 10^{12}~\si{GeV}\quad \text{with}\quad 10^{-4}<y_1<10^{-3},
\\
&& 10^{-3} < z_\mathrm{i} < 3,\quad
0 < z_\mathrm{r} < \pi/4,\quad
3 < M_{N_2}/M_{N_1} < 10^3,
\\
&& 10^{-3}~\si{GeV} < \kappa < 10^{2}~\si{GeV},\quad 
10^{-5} < \lambda_{\phi H} < 10^{-2}.
\end{eqnarray}%
Only the parameter points resulting in $\rho_\phi/\rho_\mathrm{R} < 0.1$ when $\langle\Gamma_\phi\rangle = H$ are kept for the following analysis.
In Fig.~\ref{scatter0}, we project the parameter points into the $M_{N_1}$-$\eta_B$ plane with colors indicating the $CP$ asymmetry $\varepsilon_1$.
It clearly shows that $\eta_B$ is positively correlated to both $M_{N_1}$ and $\varepsilon_1$.
In the standard leptogenesis, the observed baryon-to-photon ratio, which is denoted by the horizontal line, can be achieved for $M_{N_1} \gtrsim \num{3e10}~\si{GeV}$.
On the other hand, as the scalar-assisted leptogenesis provides an extra nonthermal source for $N_1$ particles, the applicable value of $M_{N_1}$ can be lowered down to $\sim \num{4e9}~\si{GeV}$ for the correct $\eta_B$.
Besides, the required values of $\varepsilon_1$ in the scalar-assisted case are smaller than the standard case.

\begin{figure}[!t]
\centering
\subfigure{\includegraphics[width=0.48\textwidth]{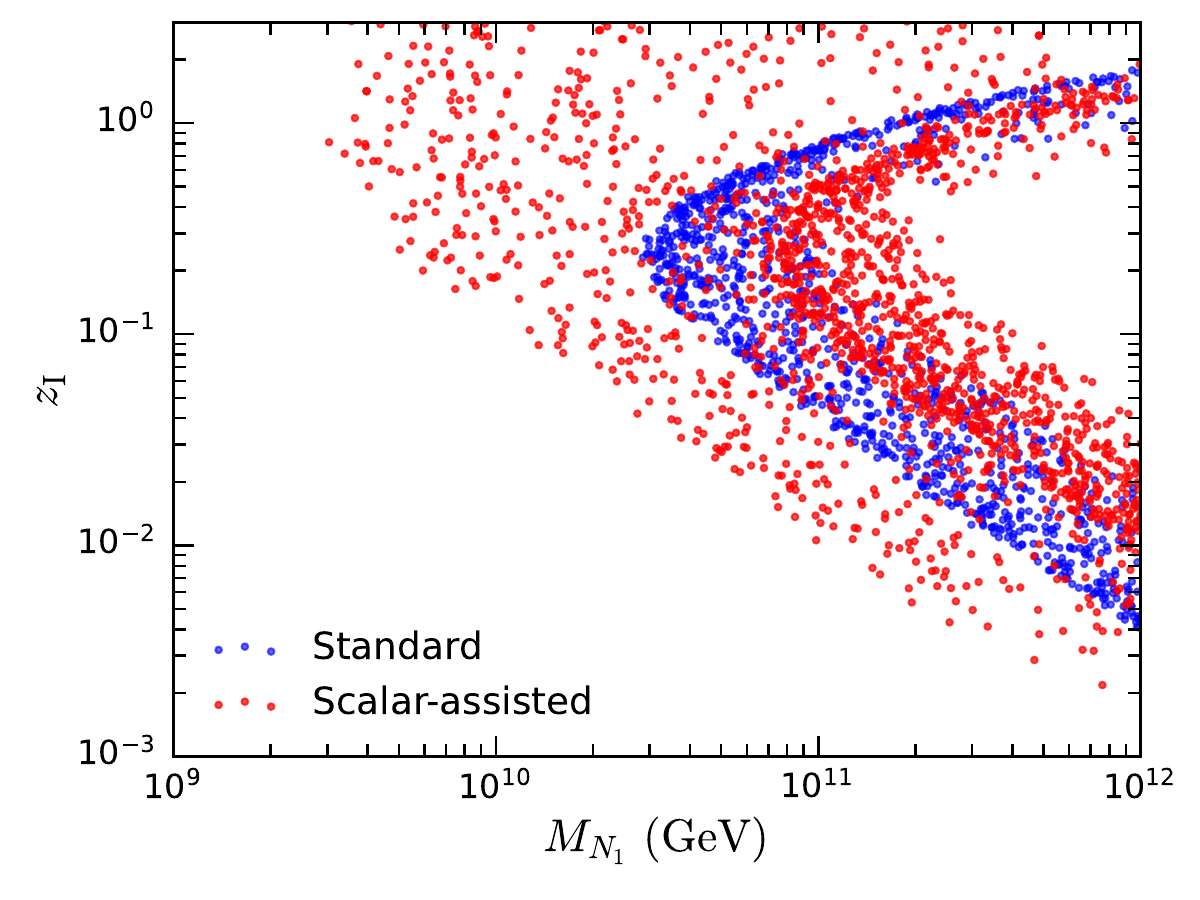}}
\hspace{.01\textwidth}
\subfigure{\includegraphics[width=0.48\textwidth]{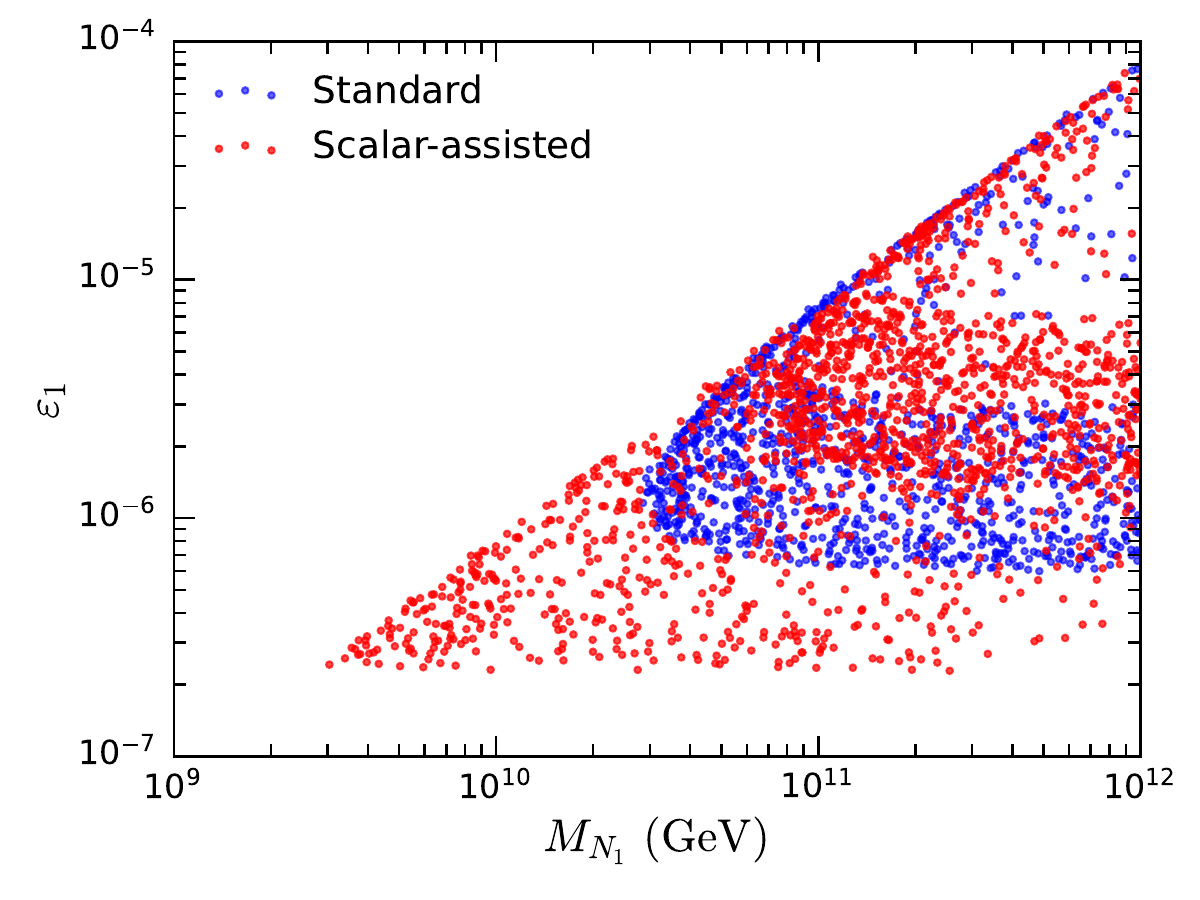}}
\caption{Selected parameter points with the correct baryon-to-photon ratio projected into the $M_{N_1}$-$z_\mathrm{i}$ (left) and $M_{N_1}$-$\varepsilon_1$ (right) planes.
The red and blue dots correspond to the standard and scalar-assisted leptogenesis, respectively.}
\label{MCcontour}
\end{figure}

In order to demonstrate the viable parameter ranges, we pick out the parameter points that lead to a baryon-to-photon ratio $\eta_B$ within the $3\sigma$ range of the observational value \eqref{eq:eta_B_obs}.
These parameter points are projected into the $M_{N_1}$-$z_\mathrm{i}$ and $M_{N_1}$-$\varepsilon_1$ planes, as illustrated in the left and right panels of Fig.~\ref{MCcontour}, respectively.
We find that the introduction of the scalar $\phi$ significantly enlarges the available parameter regions.
Compared to the standard leptogenesis, the applicable $M_{N_1}$ for the same $z_\mathrm{i}$ can be lowered down by one to three orders of magnitude.
The lowest viable $M_{N_1}$ is reduced from $\sim\num{3e10}~\si{GeV}$ to $\sim\num{3e9}~\si{GeV}$, while the smallest viable $\varepsilon_1$ decreases from $\sim \num{6e-7}$ to $\sim \num{2e-7}$.

\section{Summary}
\label{sec:sum}

In order to achieve the observed baryon asymmetry, the mass scale of RHNs in the standard leptogenesis should be rather high to provide a sufficient $CP$ violation and to overcome the washout effect.
In this work, we have proposed a particular scenario to lower down the RHN mass scale by introducing a scalar $\phi$ decaying into RHNs because of its Yukawa couplings.
$\phi$ particles should have an adequate long lifetime so that their decays provide an additional nonthermal source for the RHNs after the washout effect is attenuated.
Therefore, such a scalar-assisted leptogenesis can improve the generation of the lepton asymmetry, and the required RHN mass scale would be lower than the standard case.

To investigate the efficiency of our proposal, we have conducted a detailed analysis based on the one-flavor approximation, where only the $N_1 \leftrightarrow \ell H/\bar{\ell} \bar{H}$ processes are relevant to the generation of the lepton asymmetry.
A set of Boltzmann equations describing the evolution of the related number densities are established and solved.
The number densities as functions of $x = M_{N_1}/T$ are demonstrated for two BPs.
For BP~I with a strong washout effect, we have found that the inclusion of the scalar $\phi$ can lift up the final baryon-to-photon ratio $\eta_B$ by four orders of magnitude, compared to the standard leptogenesis.
For BP~II with a weak washout effect, $\eta_B$ can be increased by one order of magnitude in the scalar-assisted leptogenesis.

Furthermore, we have carried out parameter scans to explore the parameter space.
A scan in the $M_{N_1}$-$z_\mathrm{i}$ plane with other parameters fixed shows that the resulting baryon asymmetry is widely increased in the scalar-assisted leptogenesis, especially in the parameter regions where the washout effect is strong.
At $z_\mathrm{i} \sim 1$, which leads to a very strong washout, the lightest RHN mass $M_{N_1}$ required for the observed $\eta_B$ can be lowered down by nearly two orders of magnitude.
Based on a random scan in the parameter space, we have demonstrated the positive correlations of $\eta_B$ to $M_{N_1}$ and $\varepsilon_1$, and that both the viable $M_{N_1}$ and $\varepsilon_1$ in the scalar-assisted case are smaller than those in the standard case.
Moreover, from the parameter points that are consistent with the observed baryon-to-photon ratio in the $3\sigma$ range, we have found that the scalar-assisted leptogenesis can effectively extend the available parameter ranges, lowering down the viable $M_{N_1}$ by one to three orders of magnitude for the same $z_\mathrm{i}$.

These results clearly show that the scalar-assisted leptogenesis is a feasible avenue to lower down the RHN mass scale, particularly when the washout effect is strong.
A lower mass scale may more easily be related to rich phenomena that can be tested in experiments.
Potential implications of our work may extend beyond the specific model investigated here, offering valuable insights into the leptogenesis mechanism in diverse theoretical frameworks.
Since there are a lot of extra scalar fields in new physics beyond the SM, the origin of the scalar $\phi$ could be linked to other interesting topics, such as the origin of dark matter, grand unification, supersymmetry, and so on.

	\begin{acknowledgments}
		
		This work is supported by the National Natural Science Foundation of China (NSFC) under Grants No.~12275367 and No.~11875327, the Fundamental Research Funds for the Central Universities, the Guangzhou Science and Technology Planning Project under Grants No.~2024A04J4026, and the Sun Yat-Sen University Science Foundation.
		
	\end{acknowledgments}

	\bibliographystyle{utphys}
	
	\bibliography{ref1}

\providecommand{\href}[2]{#2}\begingroup\raggedright\begin{thebibliography}{10}

\bibitem{Cohen:1997ac}
A.~G. Cohen, A.~De~Rujula, and S.~L. Glashow, ``{A Matter - antimatter
  universe?},'' \href{http://dx.doi.org/10.1086/305328}{{\em Astrophys. J.}
  {\bfseries 495} (1998) 539--549},
  \href{http://arxiv.org/abs/astro-ph/9707087}{{\ttfamily
  arXiv:astro-ph/9707087}}.

\bibitem{Planck:2018nkj}
{\bfseries Planck} Collaboration, N.~Aghanim {\em et~al.}, ``{Planck 2018
  results. I. Overview and the cosmological legacy of Planck},''
  \href{http://dx.doi.org/10.1051/0004-6361/201833880}{{\em Astron. Astrophys.}
  {\bfseries 641} (2020) A1}, \href{http://arxiv.org/abs/1807.06205}{{\ttfamily
  arXiv:1807.06205 [astro-ph.CO]}}.

\bibitem{Fields:2019pfx}
B.~D. Fields, K.~A. Olive, T.-H. Yeh, and C.~Young, ``{Big-Bang Nucleosynthesis
  after Planck},'' \href{http://dx.doi.org/10.1088/1475-7516/2020/03/010}{{\em
  JCAP} {\bfseries 03} (2020) 010},
  \href{http://arxiv.org/abs/1912.01132}{{\ttfamily arXiv:1912.01132
  [astro-ph.CO]}}. [Erratum: JCAP 11, E02 (2020)].

\bibitem{Sakharov:1967dj}
A.~D. Sakharov, ``{Violation of CP Invariance, C asymmetry, and baryon
  asymmetry of the universe},''
  \href{http://dx.doi.org/10.1070/PU1991v034n05ABEH002497}{{\em Pisma Zh. Eksp.
  Teor. Fiz.} {\bfseries 5} (1967) 32--35}.

\bibitem{tHooft:1976rip}
G.~'t~Hooft, ``{Symmetry Breaking Through Bell-Jackiw Anomalies},''
  \href{http://dx.doi.org/10.1103/PhysRevLett.37.8}{{\em Phys. Rev. Lett.}
  {\bfseries 37} (1976) 8--11}.

\bibitem{Rubakov:1996vz}
V.~A. Rubakov and M.~E. Shaposhnikov, ``{Electroweak baryon number
  nonconservation in the early universe and in high-energy collisions},''
  \href{http://dx.doi.org/10.1070/PU1996v039n05ABEH000145}{{\em Usp. Fiz. Nauk}
  {\bfseries 166} (1996) 493--537},
  \href{http://arxiv.org/abs/hep-ph/9603208}{{\ttfamily arXiv:hep-ph/9603208}}.

\bibitem{Kuzmin:1985mm}
V.~A. Kuzmin, V.~A. Rubakov, and M.~E. Shaposhnikov, ``{On the Anomalous
  Electroweak Baryon Number Nonconservation in the Early Universe},''
  \href{http://dx.doi.org/10.1016/0370-2693(85)91028-7}{{\em Phys. Lett. B}
  {\bfseries 155} (1985) 36}.

\bibitem{Buchmuller:1994qy}
W.~Buchmuller and O.~Philipsen, ``{Phase structure and phase transition of the
  SU(2) Higgs model in three-dimensions},''
  \href{http://dx.doi.org/10.1016/0550-3213(95)00124-B}{{\em Nucl. Phys. B}
  {\bfseries 443} (1995) 47--69},
  \href{http://arxiv.org/abs/hep-ph/9411334}{{\ttfamily arXiv:hep-ph/9411334}}.

\bibitem{Kajantie:1996mn}
K.~Kajantie, M.~Laine, K.~Rummukainen, and M.~E. Shaposhnikov, ``{Is there a~
  hot electroweak phase transition at $m_H \gtrsim m_W$?},''
  \href{http://dx.doi.org/10.1103/PhysRevLett.77.2887}{{\em Phys. Rev. Lett.}
  {\bfseries 77} (1996) 2887--2890},
  \href{http://arxiv.org/abs/hep-ph/9605288}{{\ttfamily arXiv:hep-ph/9605288}}.

\bibitem{Csikor:1998eu}
F.~Csikor, Z.~Fodor, and J.~Heitger, ``{Endpoint of the hot electroweak phase
  transition},'' \href{http://dx.doi.org/10.1103/PhysRevLett.82.21}{{\em Phys.
  Rev. Lett.} {\bfseries 82} (1999) 21--24},
  \href{http://arxiv.org/abs/hep-ph/9809291}{{\ttfamily arXiv:hep-ph/9809291}}.

\bibitem{Dine:2003ax}
M.~Dine and A.~Kusenko, ``{The Origin of the matter - antimatter asymmetry},''
  \href{http://dx.doi.org/10.1103/RevModPhys.76.1}{{\em Rev. Mod. Phys.}
  {\bfseries 76} (2003) 1},
  \href{http://arxiv.org/abs/hep-ph/0303065}{{\ttfamily arXiv:hep-ph/0303065}}.

\bibitem{Bodeker:2020ghk}
D.~Bodeker and W.~Buchmuller, ``{Baryogenesis from the weak scale to the grand
  unification scale},''
  \href{http://dx.doi.org/10.1103/RevModPhys.93.035004}{{\em Rev. Mod. Phys.}
  {\bfseries 93} (2021) 035004},
  \href{http://arxiv.org/abs/2009.07294}{{\ttfamily arXiv:2009.07294
  [hep-ph]}}.

\bibitem{Pereira:2023xiw}
D.~S. Pereira, J.~a. Ferraz, F.~S.~N. Lobo, and J.~P. Mimoso, ``{Baryogenesis:
  A Symmetry Breaking in the Primordial Universe Revisited},''
  \href{http://dx.doi.org/10.3390/sym16010013}{{\em Symmetry} {\bfseries 16}
  (2024) 13}, \href{http://arxiv.org/abs/2312.14080}{{\ttfamily
  arXiv:2312.14080 [gr-qc]}}.

\bibitem{Fukugita:1986hr}
M.~Fukugita and T.~Yanagida, ``{Baryogenesis Without Grand Unification},''
  \href{http://dx.doi.org/10.1016/0370-2693(86)91126-3}{{\em Phys. Lett. B}
  {\bfseries 174} (1986) 45--47}.

\bibitem{Luty:1992un}
M.~A. Luty, ``{Baryogenesis via leptogenesis},''
  \href{http://dx.doi.org/10.1103/PhysRevD.45.455}{{\em Phys. Rev. D}
  {\bfseries 45} (1992) 455--465}.

\bibitem{Minkowski:1977sc}
P.~Minkowski, ``{$\mu \to e\gamma$ at a Rate of One Out of $10^{9}$ Muon
  Decays?},'' \href{http://dx.doi.org/10.1016/0370-2693(77)90435-X}{{\em Phys.
  Lett. B} {\bfseries 67} (1977) 421--428}.

\bibitem{Gell-Mann:1979vob}
M.~Gell-Mann, P.~Ramond, and R.~Slansky, ``{Complex Spinors and Unified
  Theories},'' {\em Conf. Proc. C} {\bfseries 790927} (1979) 315--321,
  \href{http://arxiv.org/abs/1306.4669}{{\ttfamily arXiv:1306.4669 [hep-th]}}.

\bibitem{Yanagida:1979as}
T.~Yanagida, ``{Horizontal gauge symmetry and masses of neutrinos},'' {\em
  Conf. Proc. C} {\bfseries 7902131} (1979) 95--99.

\bibitem{Kusenko:2014uta}
A.~Kusenko, K.~Schmitz, and T.~T. Yanagida, ``{Leptogenesis via Axion
  Oscillations after Inflation},''
  \href{http://dx.doi.org/10.1103/PhysRevLett.115.011302}{{\em Phys. Rev.
  Lett.} {\bfseries 115} (2015) 011302},
  \href{http://arxiv.org/abs/1412.2043}{{\ttfamily arXiv:1412.2043 [hep-ph]}}.

\bibitem{Pearce:2015nga}
L.~Pearce, L.~Yang, A.~Kusenko, and M.~Peloso, ``{Leptogenesis via neutrino
  production during Higgs condensate relaxation},''
  \href{http://dx.doi.org/10.1103/PhysRevD.92.023509}{{\em Phys. Rev. D}
  {\bfseries 92} (2015) 023509},
  \href{http://arxiv.org/abs/1505.02461}{{\ttfamily arXiv:1505.02461
  [hep-ph]}}.

\bibitem{Pascoli:2016gkf}
S.~Pascoli, J.~Turner, and Y.-L. Zhou, ``{Baryogenesis via leptonic
  CP-violating phase transition},''
  \href{http://dx.doi.org/10.1016/j.physletb.2018.03.011}{{\em Phys. Lett. B}
  {\bfseries 780} (2018) 313--318},
  \href{http://arxiv.org/abs/1609.07969}{{\ttfamily arXiv:1609.07969
  [hep-ph]}}.

\bibitem{Agashe:2018oyk}
K.~Agashe, P.~Du, M.~Ekhterachian, C.~S. Fong, S.~Hong, and L.~Vecchi,
  ``{Hybrid seesaw leptogenesis and TeV singlets},''
  \href{http://dx.doi.org/10.1016/j.physletb.2018.09.006}{{\em Phys. Lett. B}
  {\bfseries 785} (2018) 489--497},
  \href{http://arxiv.org/abs/1804.06847}{{\ttfamily arXiv:1804.06847
  [hep-ph]}}.

\bibitem{Barrie:2021mwi}
N.~D. Barrie, C.~Han, and H.~Murayama, ``{Affleck-Dine Leptogenesis from Higgs
  Inflation},'' \href{http://dx.doi.org/10.1103/PhysRevLett.128.141801}{{\em
  Phys. Rev. Lett.} {\bfseries 128} (2022) 141801},
  \href{http://arxiv.org/abs/2106.03381}{{\ttfamily arXiv:2106.03381
  [hep-ph]}}.

\bibitem{Berman:2022oht}
J.~Berman, B.~Shuve, and D.~Tucker-Smith, ``{Freeze-in leptogenesis via
  dark-matter oscillations},''
  \href{http://dx.doi.org/10.1103/PhysRevD.105.095027}{{\em Phys. Rev. D}
  {\bfseries 105} (2022) 095027},
  \href{http://arxiv.org/abs/2201.11502}{{\ttfamily arXiv:2201.11502
  [hep-ph]}}.

\bibitem{Bhattacharya:2021jli}
S.~Bhattacharya, R.~Roshan, A.~Sil, and D.~Vatsyayan, ``{Symmetry origin of
  baryon asymmetry, dark matter, and neutrino mass},''
  \href{http://dx.doi.org/10.1103/PhysRevD.106.075005}{{\em Phys. Rev. D}
  {\bfseries 106} (2022) 075005},
  \href{http://arxiv.org/abs/2105.06189}{{\ttfamily arXiv:2105.06189
  [hep-ph]}}.

\bibitem{Fernandez-Martinez:2022gsu}
E.~Fern\'andez-Mart\'\i{}nez, X.~Marcano, and D.~Naredo-Tuero, ``{HNL mass
  degeneracy: implications for low-scale seesaws, LNV at colliders and
  leptogenesis},'' \href{http://dx.doi.org/10.1007/JHEP03(2023)057}{{\em JHEP}
  {\bfseries 03} (2023) 057}, \href{http://arxiv.org/abs/2209.04461}{{\ttfamily
  arXiv:2209.04461 [hep-ph]}}.

\bibitem{Carrasco-Martinez:2023nit}
J.~Carrasco-Martinez, D.~I. Dunsky, L.~J. Hall, and K.~Harigaya,
  ``{Leptogenesis in Parity Solutions to the Strong CP Problem and Standard
  Model Parameters},'' \href{http://arxiv.org/abs/2307.15731}{{\ttfamily
  arXiv:2307.15731 [hep-ph]}}.

\bibitem{Datta:2021gyi}
A.~Datta, R.~Roshan, and A.~Sil, ``{Scalar triplet flavor leptogenesis with
  dark matter},'' \href{http://dx.doi.org/10.1103/PhysRevD.105.095032}{{\em
  Phys. Rev. D} {\bfseries 105} (2022) 095032},
  \href{http://arxiv.org/abs/2110.03914}{{\ttfamily arXiv:2110.03914
  [hep-ph]}}.

\bibitem{Co:2019wyp}
R.~T. Co and K.~Harigaya, ``{Axiogenesis},''
  \href{http://dx.doi.org/10.1103/PhysRevLett.124.111602}{{\em Phys. Rev.
  Lett.} {\bfseries 124} (2020) 111602},
  \href{http://arxiv.org/abs/1910.02080}{{\ttfamily arXiv:1910.02080
  [hep-ph]}}.

\bibitem{Zhao:2020bzx}
Z.-h. Zhao, ``{Renormalization group evolution induced leptogenesis in the
  minimal seesaw model with the trimaximal mixing and mu-tau reflection
  symmetry},'' \href{http://dx.doi.org/10.1007/JHEP11(2021)170}{{\em JHEP}
  {\bfseries 11} (2021) 170}, \href{http://arxiv.org/abs/2003.00654}{{\ttfamily
  arXiv:2003.00654 [hep-ph]}}.

\bibitem{Enomoto:2020lpf}
S.~Enomoto, C.~Cai, Z.-H. Yu, and H.-H. Zhang, ``{Leptogenesis due to
  oscillating Higgs field},''
  \href{http://dx.doi.org/10.1140/epjc/s10052-020-08672-7}{{\em Eur. Phys. J.
  C} {\bfseries 80} (2020) 1098},
  \href{http://arxiv.org/abs/2005.08037}{{\ttfamily arXiv:2005.08037
  [hep-ph]}}.

\bibitem{Jiang:2020kbt}
X.-M. Jiang, Y.-L. Tang, Z.-H. Yu, and H.-H. Zhang, ``{$1 \leftrightarrow 2$
  processes of a sterile neutrino around the electroweak scale in a thermal
  plasma},'' \href{http://dx.doi.org/10.1103/PhysRevD.103.095003}{{\em Phys.
  Rev. D} {\bfseries 103} (2021) 095003},
  \href{http://arxiv.org/abs/2008.00642}{{\ttfamily arXiv:2008.00642
  [hep-ph]}}.

\bibitem{DuttaBanik:2020vfr}
A.~Dutta~Banik, R.~Roshan, and A.~Sil, ``{Neutrino mass and asymmetric dark
  matter: study with inert Higgs doublet and high scale validity},''
  \href{http://dx.doi.org/10.1088/1475-7516/2021/03/037}{{\em JCAP} {\bfseries
  03} (2021) 037}, \href{http://arxiv.org/abs/2011.04371}{{\ttfamily
  arXiv:2011.04371 [hep-ph]}}.

\bibitem{Granelli:2023egb}
A.~Granelli, K.~Hamaguchi, N.~Nagata, M.~E. Ramirez-Quezada, and J.~Wada,
  ``{Thermal leptogenesis in the minimal gauged $ \textrm{U}{(1)}_{L_{\mu
  }-{L}_{\tau }} $ model},''
  \href{http://dx.doi.org/10.1007/JHEP09(2023)079}{{\em JHEP} {\bfseries 09}
  (2023) 079}, \href{http://arxiv.org/abs/2305.18100}{{\ttfamily
  arXiv:2305.18100 [hep-ph]}}.

\bibitem{Chen:2024arl}
S.-L. Chen, Z.~Kang, Z.-K. Liu, and P.~Zhang, ``{Matter Asymmetries in the
  $Z_N$ Dark matter-companion Models},''
  \href{http://arxiv.org/abs/2405.05694}{{\ttfamily arXiv:2405.05694
  [hep-ph]}}.

\bibitem{Davidson:2002qv}
S.~Davidson and A.~Ibarra, ``{A Lower bound on the right-handed neutrino mass
  from leptogenesis},''
  \href{http://dx.doi.org/10.1016/S0370-2693(02)01735-5}{{\em Phys. Lett. B}
  {\bfseries 535} (2002) 25--32},
  \href{http://arxiv.org/abs/hep-ph/0202239}{{\ttfamily arXiv:hep-ph/0202239}}.

\bibitem{Buchmuller:2002rq}
W.~Buchmuller, P.~Di~Bari, and M.~Plumacher, ``{Cosmic microwave background,
  matter - antimatter asymmetry and neutrino masses},''
  \href{http://dx.doi.org/10.1016/S0550-3213(02)00737-X}{{\em Nucl. Phys. B}
  {\bfseries 643} (2002) 367--390},
  \href{http://arxiv.org/abs/hep-ph/0205349}{{\ttfamily arXiv:hep-ph/0205349}}.
  [Erratum: Nucl.Phys.B 793, 362 (2008)].

\bibitem{Buchmuller:2002jk}
W.~Buchmuller, P.~Di~Bari, and M.~Plumacher, ``{A Bound on neutrino masses from
  baryogenesis},'' \href{http://dx.doi.org/10.1016/S0370-2693(02)02758-2}{{\em
  Phys. Lett. B} {\bfseries 547} (2002) 128--132},
  \href{http://arxiv.org/abs/hep-ph/0209301}{{\ttfamily arXiv:hep-ph/0209301}}.

\bibitem{Buchmuller:2003gz}
W.~Buchmuller, P.~Di~Bari, and M.~Plumacher, ``{The Neutrino mass window for
  baryogenesis},'' \href{http://dx.doi.org/10.1016/S0550-3213(03)00449-8}{{\em
  Nucl. Phys. B} {\bfseries 665} (2003) 445--468},
  \href{http://arxiv.org/abs/hep-ph/0302092}{{\ttfamily arXiv:hep-ph/0302092}}.

\bibitem{Asaka:1999yd}
T.~Asaka, K.~Hamaguchi, M.~Kawasaki, and T.~Yanagida, ``{Leptogenesis in
  inflaton decay},''
  \href{http://dx.doi.org/10.1016/S0370-2693(99)01020-5}{{\em Phys. Lett. B}
  {\bfseries 464} (1999) 12--18},
  \href{http://arxiv.org/abs/hep-ph/9906366}{{\ttfamily arXiv:hep-ph/9906366}}.

\bibitem{Hahn-Woernle:2008tsk}
F.~Hahn-Woernle and M.~Plumacher, ``{Effects of reheating on leptogenesis},''
  \href{http://dx.doi.org/10.1016/j.nuclphysb.2008.07.032}{{\em Nucl. Phys. B}
  {\bfseries 806} (2009) 68--83},
  \href{http://arxiv.org/abs/0801.3972}{{\ttfamily arXiv:0801.3972 [hep-ph]}}.

\bibitem{Barman:2021tgt}
B.~Barman, D.~Borah, and R.~Roshan, ``{Nonthermal leptogenesis and UV freeze-in
  of dark matter: Impact of inflationary reheating},''
  \href{http://dx.doi.org/10.1103/PhysRevD.104.035022}{{\em Phys. Rev. D}
  {\bfseries 104} (2021) 035022},
  \href{http://arxiv.org/abs/2103.01675}{{\ttfamily arXiv:2103.01675
  [hep-ph]}}.

\bibitem{Domcke:2020quw}
V.~Domcke, K.~Kamada, K.~Mukaida, K.~Schmitz, and M.~Yamada, ``{Wash-In
  Leptogenesis},'' \href{http://dx.doi.org/10.1103/PhysRevLett.126.201802}{{\em
  Phys. Rev. Lett.} {\bfseries 126} (2021) 201802},
  \href{http://arxiv.org/abs/2011.09347}{{\ttfamily arXiv:2011.09347
  [hep-ph]}}.

\bibitem{Domcke:2022kfs}
V.~Domcke, K.~Kamada, K.~Mukaida, K.~Schmitz, and M.~Yamada, ``{Wash-in
  leptogenesis after axion inflation},''
  \href{http://dx.doi.org/10.1007/JHEP01(2023)053}{{\em JHEP} {\bfseries 01}
  (2023) 053}, \href{http://arxiv.org/abs/2210.06412}{{\ttfamily
  arXiv:2210.06412 [hep-ph]}}.

\bibitem{Datta:2020bht}
S.~Datta, A.~Ghosal, and R.~Samanta, ``{Baryogenesis from ultralight primordial
  black holes and strong gravitational waves from cosmic strings},''
  \href{http://dx.doi.org/10.1088/1475-7516/2021/08/021}{{\em JCAP} {\bfseries
  08} (2021) 021}, \href{http://arxiv.org/abs/2012.14981}{{\ttfamily
  arXiv:2012.14981 [hep-ph]}}.

\bibitem{Barman:2021ost}
B.~Barman, D.~Borah, S.~J. Das, and R.~Roshan, ``{Non-thermal origin of
  asymmetric dark matter from inflaton and primordial black holes},''
  \href{http://dx.doi.org/10.1088/1475-7516/2022/03/031}{{\em JCAP} {\bfseries
  03} (2022) 031}, \href{http://arxiv.org/abs/2111.08034}{{\ttfamily
  arXiv:2111.08034 [hep-ph]}}.

\bibitem{Calabrese:2023key}
R.~Calabrese, M.~Chianese, J.~Gunn, G.~Miele, S.~Morisi, and N.~Saviano,
  ``{Limits on light primordial black holes from high-scale leptogenesis},''
  \href{http://dx.doi.org/10.1103/PhysRevD.107.123537}{{\em Phys. Rev. D}
  {\bfseries 107} (2023) 123537},
  \href{http://arxiv.org/abs/2305.13369}{{\ttfamily arXiv:2305.13369
  [hep-ph]}}.

\bibitem{Schmitz:2023pfy}
K.~Schmitz and X.-J. Xu, ``{Wash-in leptogenesis after the evaporation of
  primordial black holes},'' \href{http://arxiv.org/abs/2311.01089}{{\ttfamily
  arXiv:2311.01089 [hep-ph]}}.

\bibitem{Ghoshal:2023fno}
A.~Ghoshal, Y.~F. Perez-Gonzalez, and J.~Turner, ``{Superradiant
  Leptogenesis},'' \href{http://arxiv.org/abs/2312.06768}{{\ttfamily
  arXiv:2312.06768 [hep-ph]}}.

\bibitem{Suematsu:2019kst}
D.~Suematsu, ``{Low scale leptogenesis in a hybrid model of the scotogenic type
  I and III seesaw models},''
  \href{http://dx.doi.org/10.1103/PhysRevD.100.055008}{{\em Phys. Rev. D}
  {\bfseries 100} (2019) 055008},
  \href{http://arxiv.org/abs/1906.12008}{{\ttfamily arXiv:1906.12008
  [hep-ph]}}.

\bibitem{Huang:2023gse}
P.~Huang and T.~Xu, ``{Leptogenesis with a Coupling Knob},''
  \href{http://arxiv.org/abs/2312.06380}{{\ttfamily arXiv:2312.06380
  [hep-ph]}}.

\bibitem{Chun:2023ezg}
E.~J. Chun, T.~P. Dutka, T.~H. Jung, X.~Nagels, and M.~Vanvlasselaer,
  ``{Bubble-assisted leptogenesis},''
  \href{http://dx.doi.org/10.1007/JHEP09(2023)164}{{\em JHEP} {\bfseries 09}
  (2023) 164}, \href{http://arxiv.org/abs/2305.10759}{{\ttfamily
  arXiv:2305.10759 [hep-ph]}}.

\bibitem{Dehpour:2023dfo}
M.~Dehpour, ``{Thermal leptogenesis in nonextensive cosmology},''
  \href{http://dx.doi.org/10.1140/epjc/s10052-024-12697-7}{{\em Eur. Phys. J.
  C} {\bfseries 84} (2024) 340},
  \href{http://arxiv.org/abs/2401.00229}{{\ttfamily arXiv:2401.00229
  [hep-ph]}}.

\bibitem{Dehpour:2023wyy}
M.~Dehpour, ``{Thermal leptogenesis in anisotropic cosmology},''
  \href{http://dx.doi.org/10.1142/S0217751X23501816}{{\em Int. J. Mod. Phys. A}
  {\bfseries 38} (2023) 2350181},
  \href{http://arxiv.org/abs/2312.10677}{{\ttfamily arXiv:2312.10677
  [hep-ph]}}.

\bibitem{Pilaftsis:2003gt}
A.~Pilaftsis and T.~E.~J. Underwood, ``{Resonant leptogenesis},''
  \href{http://dx.doi.org/10.1016/j.nuclphysb.2004.05.029}{{\em Nucl. Phys. B}
  {\bfseries 692} (2004) 303--345},
  \href{http://arxiv.org/abs/hep-ph/0309342}{{\ttfamily arXiv:hep-ph/0309342}}.

\bibitem{Pilaftsis:2004xx}
A.~Pilaftsis, ``{Resonant tau-leptogenesis with observable lepton number
  violation},'' \href{http://dx.doi.org/10.1103/PhysRevLett.95.081602}{{\em
  Phys. Rev. Lett.} {\bfseries 95} (2005) 081602},
  \href{http://arxiv.org/abs/hep-ph/0408103}{{\ttfamily arXiv:hep-ph/0408103}}.

\bibitem{Dev:2017wwc}
B.~Dev, M.~Garny, J.~Klaric, P.~Millington, and D.~Teresi, ``{Resonant
  enhancement in leptogenesis},''
  \href{http://dx.doi.org/10.1142/S0217751X18420034}{{\em Int. J. Mod. Phys. A}
  {\bfseries 33} (2018) 1842003},
  \href{http://arxiv.org/abs/1711.02863}{{\ttfamily arXiv:1711.02863
  [hep-ph]}}.

\bibitem{daSilva:2022mrx}
P.~C. da~Silva, D.~Karamitros, T.~McKelvey, and A.~Pilaftsis, ``{Tri-resonant
  leptogenesis in a seesaw extension of the Standard Model},''
  \href{http://dx.doi.org/10.1007/JHEP11(2022)065}{{\em JHEP} {\bfseries 11}
  (2022) 065}, \href{http://arxiv.org/abs/2206.08352}{{\ttfamily
  arXiv:2206.08352 [hep-ph]}}.

\bibitem{Karamitros:2023tqr}
D.~Karamitros, T.~McKelvey, and A.~Pilaftsis, ``{Varying entropy degrees of
  freedom effects in low-scale leptogenesis},''
  \href{http://dx.doi.org/10.1103/PhysRevD.109.055007}{{\em Phys. Rev. D}
  {\bfseries 109} (2024) 055007},
  \href{http://arxiv.org/abs/2310.03703}{{\ttfamily arXiv:2310.03703
  [hep-ph]}}.

\bibitem{Zhao:2024uid}
Z.-h. Zhao, J.~Zhang, and X.-Y. Wu, ``{Flavored leptogenesis from a sudden mass
  gain of right-handed neutrinos},''
  \href{http://arxiv.org/abs/2403.18630}{{\ttfamily arXiv:2403.18630
  [hep-ph]}}.

\bibitem{Barbieri:1999ma}
R.~Barbieri, P.~Creminelli, A.~Strumia, and N.~Tetradis, ``{Baryogenesis
  through leptogenesis},''
  \href{http://dx.doi.org/10.1016/S0550-3213(00)00011-0}{{\em Nucl. Phys. B}
  {\bfseries 575} (2000) 61--77},
  \href{http://arxiv.org/abs/hep-ph/9911315}{{\ttfamily arXiv:hep-ph/9911315}}.

\bibitem{Abada:2006ea}
A.~Abada, S.~Davidson, A.~Ibarra, F.~X. Josse-Michaux, M.~Losada, and
  A.~Riotto, ``{Flavour Matters in Leptogenesis},''
  \href{http://dx.doi.org/10.1088/1126-6708/2006/09/010}{{\em JHEP} {\bfseries
  09} (2006) 010}, \href{http://arxiv.org/abs/hep-ph/0605281}{{\ttfamily
  arXiv:hep-ph/0605281}}.

\bibitem{Moffat:2018wke}
K.~Moffat, S.~Pascoli, S.~T. Petcov, H.~Schulz, and J.~Turner,
  ``{Three-flavored nonresonant leptogenesis at intermediate scales},''
  \href{http://dx.doi.org/10.1103/PhysRevD.98.015036}{{\em Phys. Rev. D}
  {\bfseries 98} (2018) 015036},
  \href{http://arxiv.org/abs/1804.05066}{{\ttfamily arXiv:1804.05066
  [hep-ph]}}.

\bibitem{Alanne:2018brf}
T.~Alanne, T.~Hugle, M.~Platscher, and K.~Schmitz, ``{Low-scale leptogenesis
  assisted by a real scalar singlet},''
  \href{http://dx.doi.org/10.1088/1475-7516/2019/03/037}{{\em JCAP} {\bfseries
  03} (2019) 037}, \href{http://arxiv.org/abs/1812.04421}{{\ttfamily
  arXiv:1812.04421 [hep-ph]}}.

\bibitem{Dev:2017xry}
P.~S.~B. Dev, R.~N. Mohapatra, and Y.~Zhang, ``{Leptogenesis constraints on $B
  - L$ breaking Higgs boson in TeV scale seesaw models},''
  \href{http://dx.doi.org/10.1007/JHEP03(2018)122}{{\em JHEP} {\bfseries 03}
  (2018) 122}, \href{http://arxiv.org/abs/1711.07634}{{\ttfamily
  arXiv:1711.07634 [hep-ph]}}.

\bibitem{Barreiros:2022fpi}
D.~M. Barreiros, H.~B. C\^amara, R.~G. Felipe, and F.~R. Joaquim,
  ``{Scalar-singlet assisted leptogenesis with CP violation from the vacuum},''
  \href{http://dx.doi.org/10.1007/JHEP01(2023)010}{{\em JHEP} {\bfseries 01}
  (2023) 010}, \href{http://arxiv.org/abs/2211.00042}{{\ttfamily
  arXiv:2211.00042 [hep-ph]}}.

\bibitem{Casas:2001sr}
J.~A. Casas and A.~Ibarra, ``{Oscillating neutrinos and $\mu \to e, \gamma$},''
  \href{http://dx.doi.org/10.1016/S0550-3213(01)00475-8}{{\em Nucl. Phys. B}
  {\bfseries 618} (2001) 171--204},
  \href{http://arxiv.org/abs/hep-ph/0103065}{{\ttfamily arXiv:hep-ph/0103065}}.

\bibitem{Pontecorvo:1957qd}
B.~Pontecorvo, ``{Inverse beta processes and nonconservation of lepton
  charge},'' {\em Zh. Eksp. Teor. Fiz.} {\bfseries 34} (1957) 247.

\bibitem{Maki:1962mu}
Z.~Maki, M.~Nakagawa, and S.~Sakata, ``{Remarks on the unified model of
  elementary particles},'' \href{http://dx.doi.org/10.1143/PTP.28.870}{{\em
  Prog. Theor. Phys.} {\bfseries 28} (1962) 870--880}.

\bibitem{Workman:2022ynf}
{\bfseries Particle Data Group} Collaboration, R.~L. Workman and Others,
  ``{Review of Particle Physics},''
  \href{http://dx.doi.org/10.1093/ptep/ptac097}{{\em PTEP} {\bfseries 2022}
  (2022) 083C01}.

\bibitem{Gonzalez-Garcia:2021dve}
M.~C. Gonzalez-Garcia, M.~Maltoni, and T.~Schwetz, ``{NuFIT: Three-Flavour
  Global Analyses of Neutrino Oscillation Experiments},''
  \href{http://dx.doi.org/10.3390/universe7120459}{{\em Universe} {\bfseries 7}
  (2021) 459}, \href{http://arxiv.org/abs/2111.03086}{{\ttfamily
  arXiv:2111.03086 [hep-ph]}}.

\bibitem{Antusch:2011nz}
S.~Antusch, P.~Di~Bari, D.~A. Jones, and S.~F. King, ``{Leptogenesis in the Two
  Right-Handed Neutrino Model Revisited},''
  \href{http://dx.doi.org/10.1103/PhysRevD.86.023516}{{\em Phys. Rev. D}
  {\bfseries 86} (2012) 023516},
  \href{http://arxiv.org/abs/1107.6002}{{\ttfamily arXiv:1107.6002 [hep-ph]}}.

\bibitem{Covi:1996wh}
L.~Covi, E.~Roulet, and F.~Vissani, ``{CP violating decays in leptogenesis
  scenarios},'' \href{http://dx.doi.org/10.1016/0370-2693(96)00817-9}{{\em
  Phys. Lett. B} {\bfseries 384} (1996) 169--174},
  \href{http://arxiv.org/abs/hep-ph/9605319}{{\ttfamily arXiv:hep-ph/9605319}}.

\bibitem{Khlebnikov:1988sr}
S.~Y. Khlebnikov and M.~E. Shaposhnikov, ``{The Statistical Theory of Anomalous
  Fermion Number Nonconservation},''
  \href{http://dx.doi.org/10.1016/0550-3213(88)90133-2}{{\em Nucl. Phys. B}
  {\bfseries 308} (1988) 885--912}.

\bibitem{Kolb:1990vq}
E.~W. Kolb and M.~S. Turner,
  \href{http://dx.doi.org/10.1201/9780429492860}{{\em {The Early Universe}}},
  vol.~69.
\newblock 1990.

\bibitem{Xing:2011zza}
Z.-Z. Xing and S.~Zhou, \href{http://dx.doi.org/10.1007/978-3-642-17560-2}{{\em
  {Neutrinos in particle physics, astronomy and cosmology}}}.
\newblock Springer, 2011.

\bibitem{Basboll:2006yx}
A.~Basboll and S.~Hannestad, ``{Decay of heavy Majorana neutrinos using the
  full Boltzmann equation including its implications for leptogenesis},''
  \href{http://dx.doi.org/10.1088/1475-7516/2007/01/003}{{\em JCAP} {\bfseries
  01} (2007) 003}, \href{http://arxiv.org/abs/hep-ph/0609025}{{\ttfamily
  arXiv:hep-ph/0609025}}.

\bibitem{Garayoa:2009my}
J.~Garayoa, S.~Pastor, T.~Pinto, N.~Rius, and O.~Vives, ``{On the full
  Boltzmann equations for Leptogenesis},''
  \href{http://dx.doi.org/10.1088/1475-7516/2009/09/035}{{\em JCAP} {\bfseries
  09} (2009) 035}, \href{http://arxiv.org/abs/0905.4834}{{\ttfamily
  arXiv:0905.4834 [hep-ph]}}.

\bibitem{Hahn-Woernle:2009jyb}
F.~Hahn-Woernle, M.~Plumacher, and Y.~Y.~Y. Wong, ``{Full Boltzmann equations
  for leptogenesis including scattering},''
  \href{http://dx.doi.org/10.1088/1475-7516/2009/08/028}{{\em JCAP} {\bfseries
  08} (2009) 028}, \href{http://arxiv.org/abs/0907.0205}{{\ttfamily
  arXiv:0907.0205 [hep-ph]}}.

\bibitem{Hofmann:2001bi}
S.~Hofmann, D.~J. Schwarz, and H.~Stoecker, ``{Damping scales of neutralino
  cold dark matter},'' \href{http://dx.doi.org/10.1103/PhysRevD.64.083507}{{\em
  Phys. Rev. D} {\bfseries 64} (2001) 083507},
  \href{http://arxiv.org/abs/astro-ph/0104173}{{\ttfamily
  arXiv:astro-ph/0104173}}.

\bibitem{Visinelli:2015eka}
L.~Visinelli and P.~Gondolo, ``{Kinetic decoupling of WIMPs: analytic
  expressions},'' \href{http://dx.doi.org/10.1103/PhysRevD.91.083526}{{\em
  Phys. Rev. D} {\bfseries 91} (2015) 083526},
  \href{http://arxiv.org/abs/1501.02233}{{\ttfamily arXiv:1501.02233
  [astro-ph.CO]}}.

\bibitem{Cai:2021wmu}
C.~Cai and H.-H. Zhang, ``{Vector dark matter production from catalyzed
  annihilation},'' \href{http://dx.doi.org/10.1007/JHEP01(2022)099}{{\em JHEP}
  {\bfseries 01} (2022) 099}, \href{http://arxiv.org/abs/2107.13475}{{\ttfamily
  arXiv:2107.13475 [hep-ph]}}.

\bibitem{Kolb:1979qa}
E.~W. Kolb and S.~Wolfram, ``{Baryon Number Generation in the Early
  Universe},'' \href{http://dx.doi.org/10.1016/0550-3213(82)90012-8}{{\em Nucl.
  Phys. B} {\bfseries 172} (1980) 224}. [Erratum: Nucl.Phys.B 195, 542 (1982)].

\bibitem{Buchmuller:2004nz}
W.~Buchmuller, P.~Di~Bari, and M.~Plumacher, ``{Leptogenesis for
  pedestrians},'' \href{http://dx.doi.org/10.1016/j.aop.2004.02.003}{{\em
  Annals Phys.} {\bfseries 315} (2005) 305--351},
  \href{http://arxiv.org/abs/hep-ph/0401240}{{\ttfamily arXiv:hep-ph/0401240}}.

\bibitem{Gondolo:1990dk}
P.~Gondolo and G.~Gelmini, ``{Cosmic abundances of stable particles: Improved
  analysis},'' \href{http://dx.doi.org/10.1016/0550-3213(91)90438-4}{{\em Nucl.
  Phys. B} {\bfseries 360} (1991) 145--179}.

\bibitem{Plumacher:1997ru}
M.~Plumacher, ``{Baryon asymmetry, neutrino mixing and supersymmetric SO(10)
  unification},'' \href{http://dx.doi.org/10.1016/S0550-3213(98)00410-6}{{\em
  Nucl. Phys. B} {\bfseries 530} (1998) 207--246},
  \href{http://arxiv.org/abs/hep-ph/9704231}{{\ttfamily arXiv:hep-ph/9704231}}.

\bibitem{Cataldi:2024bcs}
M.~Cataldi, A.~Mariotti, F.~Sala, and M.~Vanvlasselaer, ``{ALP leptogenesis:
  non-thermal right-handed neutrinos from axions},''
  \href{http://dx.doi.org/10.1007/JHEP12(2024)125}{{\em JHEP} {\bfseries 12}
  (2024) 125}, \href{http://arxiv.org/abs/2407.01667}{{\ttfamily
  arXiv:2407.01667 [hep-ph]}}.

\end{thebibliography}\endgroup

\end{document}